\documentclass{ws-ijbc}
\usepackage{ws-rotating}     
\usepackage{amssymb,amsmath}
\usepackage{url} 
\usepackage{graphicx}
\usepackage{epstopdf}
\usepackage{xspace}

\newcommand{\Poincare}{Poincar\'e\xspace}
\begin{document}

\catchline{}{}{}{}{} 

\markboth{Vinay Kumar \textit{et. al.}}{Unpredictability of the basins of attraction}

\title{FRACTAL BASINS OF ATTRACTION IN A BINARY QUASAR MODEL}

\author{Vinay Kumar}

\address{Department of Mathematics, Zakir Husain Delhi College, \\University of Delhi, Delhi, India\\ email: vkumar@zh.du.ac.in}

\author{Pankaj Sharma}
\address{Department of Mathematics, Zakir Husain Delhi College, \\University of Delhi, Delhi, India\\
drps4m@gmail.com}
\author{Rajiv Aggarwal}
\address{Department of Mathematics,  Sri Aurobindo College,\\ University of Delhi, Delhi, India\\ \url{rajiv_agg1973@yahoo.com}}
\author{Bhavneet Kaur}
\address{Department of Mathematics, Lady Sri Ram College,\\ University of Delhi, Delhi, India\\
	bhavneet.lsr@gmail.com}
%
\maketitle

\begin{history}
\received{xxxxxxxx}
\end{history}

\begin{abstract}
The present paper investigates the binary system of quasars in the framework of the Circular Restricted Three-Body Problem. The parametric evolution of libration points, the geometry of zero-velocity curves are one of the crucial aspects of our study. The multivariate form of NR method is applied to study the basin of attraction connected with libration points. The algorithm for using the Newton-Raphson method is slightly modified in order to avoid the unnecessary delay in the convergence of initial conditions. The impact of parameters on the shape of the basin of attraction and the number of iterations needed for the convergence of initial conditions are explored.  We carry out an exhaustive (numerical) study to show the influence of these parameters on converging regions in basins of convergence. We unveil the existence of fractal structure in the basin of attraction using the method of basin entropy. In almost all cases, the existence of fractal structure is found throughout the basins of attraction.
\end{abstract}

	\section{Introduction}
    The study of the complexity of the phase space of nonlinear models in the field of space dynamics and celestial mechanics have gained considerable attention nowadays.  Mathematicians who have been working in this field for decades are trying to explore several nonlinear models. They have introduced many tools to study the phase space structure of these models. The basins of attraction (BoA)(or basin of convergence (BoC)) are one of the essential tools. Our observations took place while following many recent contributions in this field such as \cite{Zotos17}, \cite{Suraj2019} and  \cite{Zotos2018}. The study of BoA reveals the regions of initial conditions which converges smoothly towards libration points of the system under consideration. However, there are regions of initial conditions which creates an obstacle for smooth convergence. It is because of the non-linearity that existed in the equations of motion for a particular system under consideration. 
    
    The galactic model that has been considered in our work is the system of binary quasars. We have studied this model under the assumption of Circular Restricted Three-Body Problem (CRTBP). In \cite{Zotos12b} and \cite{kumar16}, one can observe the applications of the \Poincare section, the method of Lyapunov exponents and wavelet transform method to understand the phase space of this model. The first time, through this work, we have investigated the BoA in the model of spaces of galaxies. In the recent past few researchers are working in these models and trying to explore the dynamics of this model. In continuation of that, we have also deliberated the parametric evolution of libration points in the value of parameters as well as the zero-velocity curves.   
    
    However, it is not easy to establish the existence of fractal structure in general. Recently, Alvar Daza \cite{Alvar16} has introduced the concept of basin entropy to show the existence of fractals in BoA. Some applications of this method can also be seen in the work of \cite{Alvar17}, \cite{Alvar18}. The method of basin entropy is the quantification of uncertainty involved in BoA. We compute the basin entropy   $\text{S}_{\text{b}}$ and entropy along the boundaries $S_{bb}$ of the BoA. If the value of   $\text{S}_{\text{b}}$ or   $\text{S}_{\text{bb}}$ is greater than $\log 2$, then BoA or boundaries along the BoA is fractal. We choose this method to reveal the existence of fractals connected with libration points of the binary system of interacting galaxies.
    
    Further, we have noticed several numerical methods while studying the BoA. Some notable works have been observed in the work of  \cite{Zotos17}, \cite{Suraj2019} and \cite{Zotos2018}. From results established in these papers, we notice that the NR-method is efficient to study in such a phenomenon. Therefore, we have considered the multivariate NR-method to study the convergence of BoA.
    
    The present paper aims to study the BoA in the system of binary quasars. Also, we have tried to investigate the existence of fractal. We have applied the multivariate NR-method and the method of basin entropy. The algorithm for finding BoA is written in such a way that it excludes unnecessary iteration time consumed by CPU.  
    
    The organisation of the present work is as follows: In Section 2, the configuration of a binary quasar model is described. The parametric evolution of the libration points and the geometry of zero-velocity curves under the effect of parameters are included in its subsections. The Newton$-$Raphson Method and the recently developed tool basin entropy are explained in Section 3. Section 4 comprises the results and discussion and finally, we have made concluding remarks related to the observations based on numerical simulations in Section 5.     
\section{Configuration of binary quasar model}
The binary quasars is the system of interacting galaxies, which are bound together by gravitational forces. It is believed to be the product of the merger of galaxies. At present, it is known that many interacting quasars are hold in massive spiral galaxies with prominent disks \cite{Lat2007}.  To date (to the best of our knowledge) OJ287 \cite{Silli1988} is the famous close binary pair of supermassive black holes (SMBH).  Our model contains a pair of disk galaxies equipped with a dense, massive and spherically symmetric nucleus. Notably, the potential which governs the motion of a star due to the first galaxy $(G1)$ is expressed as:
\begin{equation}
\centering
\begin{split}
V_{1}(r,z)=V_{n1}(r,z)+V_{d1}(r,z)=  -\frac{\text{M}_{n1}}{\sqrt{r^2+z^2+{c_{n1}}^2}}
-\frac{\text{M}_{d1}}{\sqrt{b_{1} ^2+r^2+\left(a_{1}+ \sqrt{h_{1}^2+z^2}\right)^2}},
\end{split}
\end{equation}
where  
$ r^2=x^2+y^2$. The galaxy has mass on the disk $\text{M}_{d1}$ and the mass on the nucleus is $\text{M}_{n1}$. $a_{1}$  and $h_{1}$ are the scale length   and scale height of the disk respectively.  is The core radius of the disk halo is $b_{1}$ and $c_{n1}$ is the scale length of the nucleus.\\
Similarly, the potential accountable for the motion of the star due to G2 is expressed as 
\begin{equation}
\centering
\begin{split}
V_{2}(r,z)=V_{n2}(r,z)+ V_{d2}(r,z)= -\frac{\text{M}_{n2}}{\sqrt{r^2+z^2+c_{n2}^2}} -\frac{\text{M}_{d2}}{\sqrt{b_{2}^2+r^2+\left(a_{2}+\sqrt{\left(h_{2}\right)^2+z^2}\right)^2}},
\end{split}
\end{equation}
The galaxy has mass on the disk $\text{M}_{d2}$ and the mass on the nucleus is $\text{M}_{n2}$. $a_{2}$  and $h_{2}$ are the scale length   and scale height of the disk respectively.  is The core radius of the disk halo is $b_{2}$ and $c_{n2}$ is the scale length of the nucleus.\\

Expressions for the disk of host galaxies are taken from the work of Miyamoto-Nagai \cite{Miyamoto1975} and \cite{Zotos12b}.  We have considered the Plummer sphere to express the nuclei of both galaxies \cite{Hasan1993}.  In this model,  we study the motion of a star under the influence of the binary quasars in the framework of the CRTBP. The two galaxies rotate in circular orbits in an inertial frame OXYZ, with the origin at the centre of  mass of the system with a constant angular velocity                                                                                                                                                                                    
\begin{equation}
\Omega_{p}=\sqrt{\frac{GM_{t}}{R^3}}> 0
\end{equation}
where $\text{M}_{t}=\text{M}_{n1}+ \text{M}_{d1}+ \text{M}_{n2}+ \text{M}_{d2}$. Here, the total mass of the system is $M_{t}$ and the distance between the centers of two galaxies is  $R$.  Also, the frame which is rotating with angular velocity $\Omega_{p}$ has fixed positions at $C_{1}(x,y,z)=(x_{1},0,0)$ and $C_{2}(x,y,z)= (x_{2},0,0) $, respectively.  The distance between two interacting galaxies is such that the tidal forces are very small and can be neglected.  Therefore, the net potential accountable for the motion of a star in the Binary Quasar System is
\begin{equation}
\phi_{t}(x,y,z)= \phi_{G1}(x,y,z)+\phi_{G2}(x,y,z)+\phi_{rot}(x,y,z),
\end{equation}
where
\begin{equation}
\begin{split}
\phi_{G1}(x,y,z)=-\frac{\text{M}_{n1}}{\sqrt{r_{1} ^2+{c_{n1}}^2}}  -\frac{\text{M}_{d1}}{\sqrt{b_{1}^2+r_{a1} ^2+\left(a_{1}+ \sqrt{h_{1}^2+z^2} \right)^2}},
\end{split}
\end{equation} 
\begin{equation}
\begin{split}
\phi_{G2}(x,y,z)= -\frac{\text{M}_{n2}}{\sqrt{r_{2} ^2+c_{n2}^2}}
-\frac{\text{M}_{d2}}{\sqrt{b_{2}^2+r_{a2} ^2+\left(a_{2}+\sqrt{h_{2} ^2+z^2}\right) }}, 
\end{split}
\end{equation}  
\begin{align}
\phi_{rot}(x,y) = -\frac{\Omega_{p}^2}{2}\left(\frac{\text{M}_{2}}{\text{M}_{t}} r_{a2}^2+R_{s}r_{a1}^2-R^{2}\frac{\text{M}_{2}}{\text{M}_{t}}R_{s}\right)
\end{align}

\begin{equation}\nonumber
(r_{a1})^2 = (x-x_{1})^2 + y^2, \ (r_{a2})^2 = (x-x_{2})^2 + y^2, \    (r_{1})^2 = (r_{a1})^2 + z^2, \ (r_{2})^2 = (r_{a2})^2 + z^2,
\end{equation}

\begin{equation}\nonumber
\ x_{1} = -\frac{\text{M}_{2}}{\text{M}_{t}}R,  \ x_{2} = R-\frac{\text{M}_{2}}{\text{M}_{t}},\ \text{M}_{2} = \text{M}_{n2}+\text{M}_{d2}, \ R_{s} = 1-\frac{\text{M}_{2}}{\text{M}_{t}}.
\end{equation}
 In synodic frame the equations of motions (two dimension)are,
\begin{equation}
\ddot{x}=-\frac{\partial\phi_{t}}{\partial x}-2\Omega_{p}\dot{y} \quad
\ddot{y}=-\frac{\partial\phi_{t}}{\partial y}-2\Omega_{p}\dot{x}.  
\end{equation}
The Jacobi integral for the system of equations of motion (8) is given by the equation 
\begin{equation}
J=\frac{1}{2}\left(\dot{x}^2+\dot{y}^2\right)+\phi_{t}(x,y)=E_{j},
\end{equation} 
where $\dot{x}$, $\dot{y}$ and $\dot{z}$ are the momenta corresponding to coordinates $x, y \text{and} z$ respectively.
We use the following galactic units:
\begin{itemlist}
	\item Unit of length is 20 kiloparsec (one parsec $ \simeq $ 3.26 light years).
	\item Unit of mass is $1.8 \times10^{11}$ $\text{M}$$\odot$ ($\text{M}$$\odot$ denotes the solar mass which is equal to $1.98892 \times 10^{30} kg$).
	\item Unit of time is $ 0.99 \times10^8$ year.
	\item Unit of velocity is 197 km/sec.
	\item G=1.
\end{itemlist}

In these units, we use $a_{1}=0.15, b_{1}=0.2542, h_{1}=0.00925, c_{n1}=0.0125, c_{n1}=0.0125, a_{2}=0.175, b_{2}=0.0789, h_{2}=0.00875$ and $c_{n2}=0.01$, which remain constant throughout the computations. The values of $\text{M}_{n1}, \text{M}_{d1},\text{M}_{n2}, \text{M}_{d2}$ and $R$ are  parameters. \cite{Zotos12b}
\subsection{Influence on libration points due to variation in parameters}

\begin{figure*}[htb!]
	\centering
	\includegraphics[width=2.5 in, height=2.5 in]{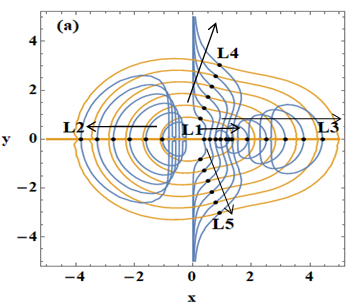}\hspace{1.5cm}
	\includegraphics[width=2.5 in, height=2.5 in]{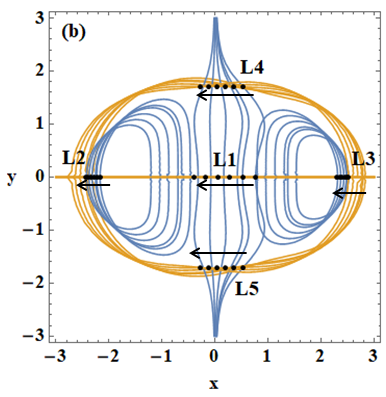}
	\caption{(a) Movement of libration points due to variation in $R$, (b) Movement of libration points due to variation in $\text{M}_{n1}$, $\text{M}_{d1}$, $\text{M}_{n2}$ and $\text{M}_{d2}$.}
\end{figure*}
In Fig.1(a) and Fig.1(b), we have presented the effect of parameter $R$ and effect of parameters $\text{M}_{n1}, \text{M}_{d1}, \text{M}_{n2}$, $\text{M}_{d2}$ on the positions of libration points, respectively. In Fig.1(a), we notice that the libration points are moving away from the origin as the value of $R$ increases in the interval [1, 3.5]. We find that $\text{L1}$ has less movement compared to $\text{L4, L5}$. The displacement of $\text{L2, L3}$ is larger than $\text{L1, L3 and L4}$. In Fig. 1(b), we are decreasing the mass of galaxy G1 and increasing the mass of galaxy G2. We observe shifts in all libration points towards the negative direction of the $x$-axis. The shift of $\text{L1}$ is comparatively more than the shift of $\text{L2, L3, L4 and L5}$. Thus the parameters have a considerable impact on the position of libration points.

\subsection{Zero-velocity curves}
\begin{figure*}[htb!]
	\centering
	\begin{tabular}{ccc}
		\includegraphics[scale=.4]{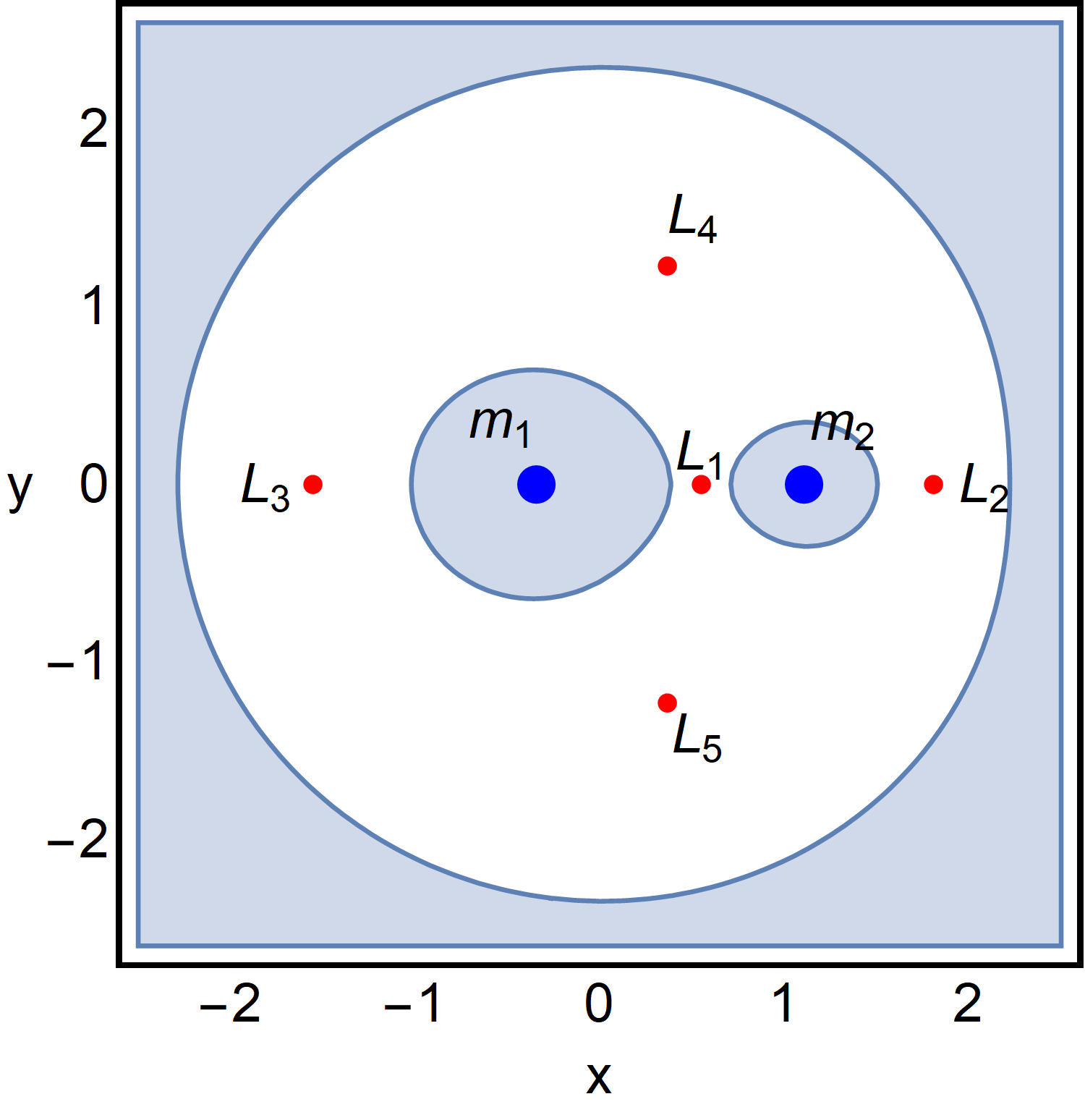}&
		\includegraphics[scale=.4]{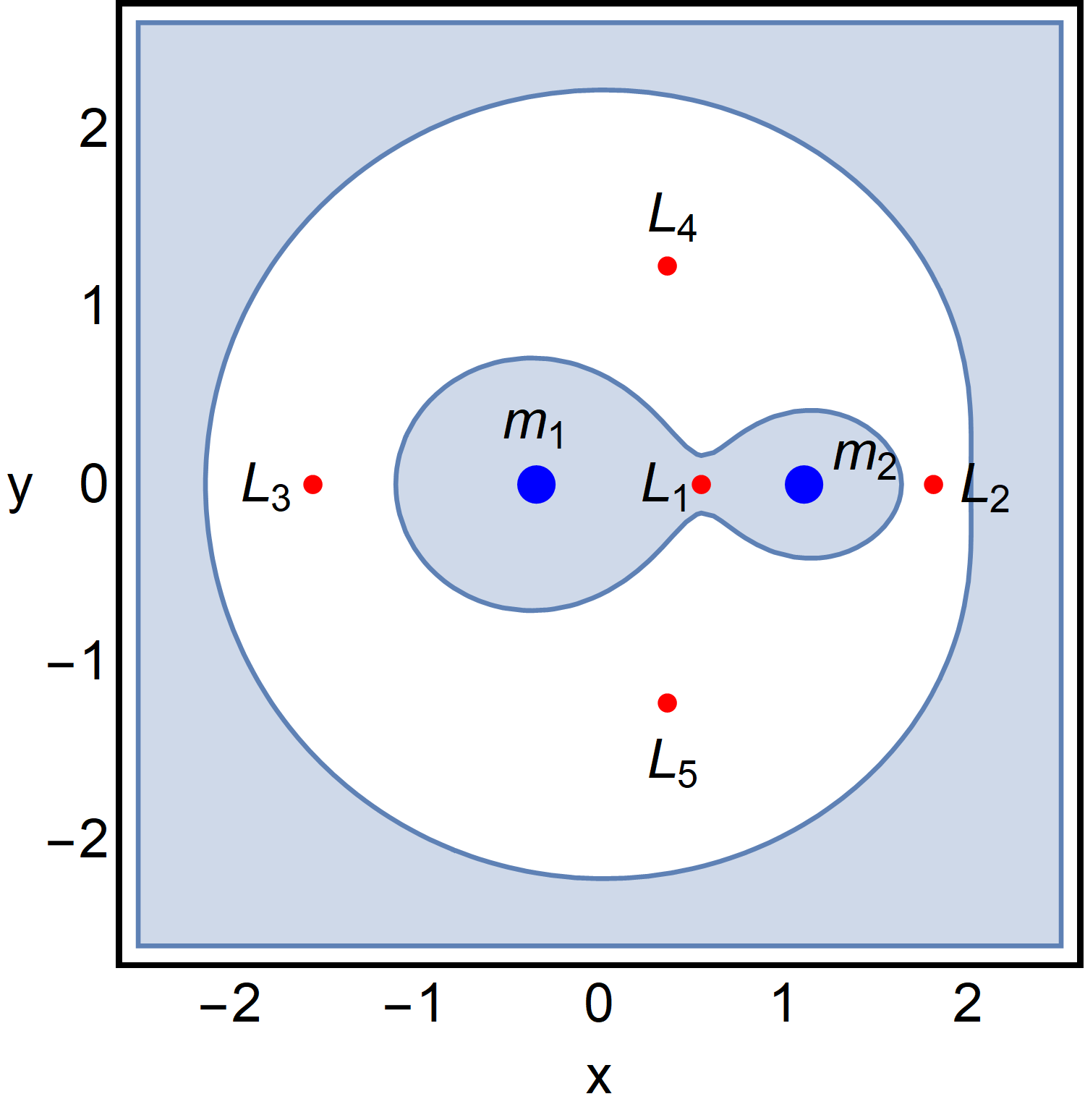}&
		\includegraphics[scale=.4]{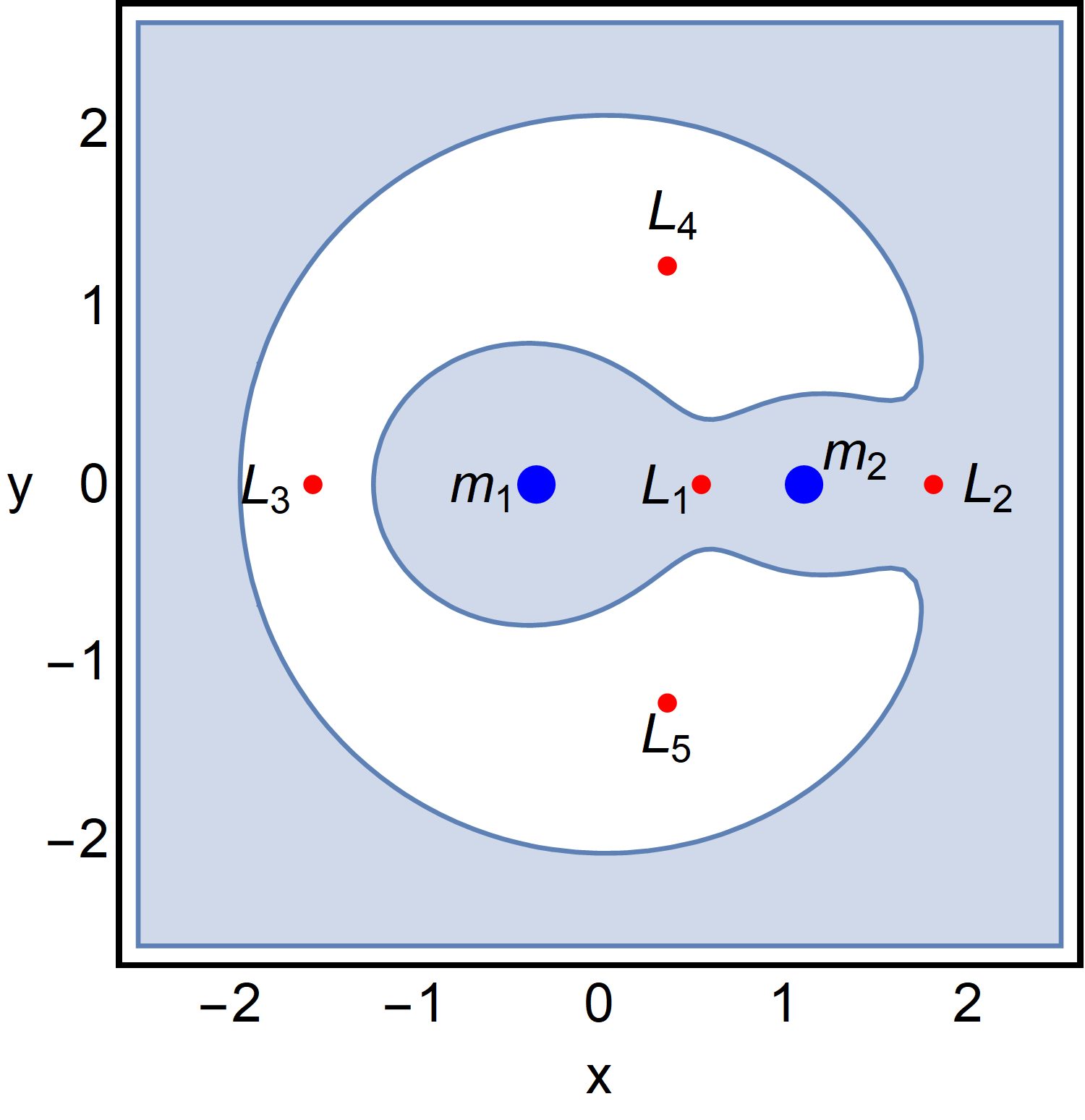}\\
		(a) C=7.15 &(b) C=6.75 &(c) C=6.35\\
		\includegraphics[scale=.4]{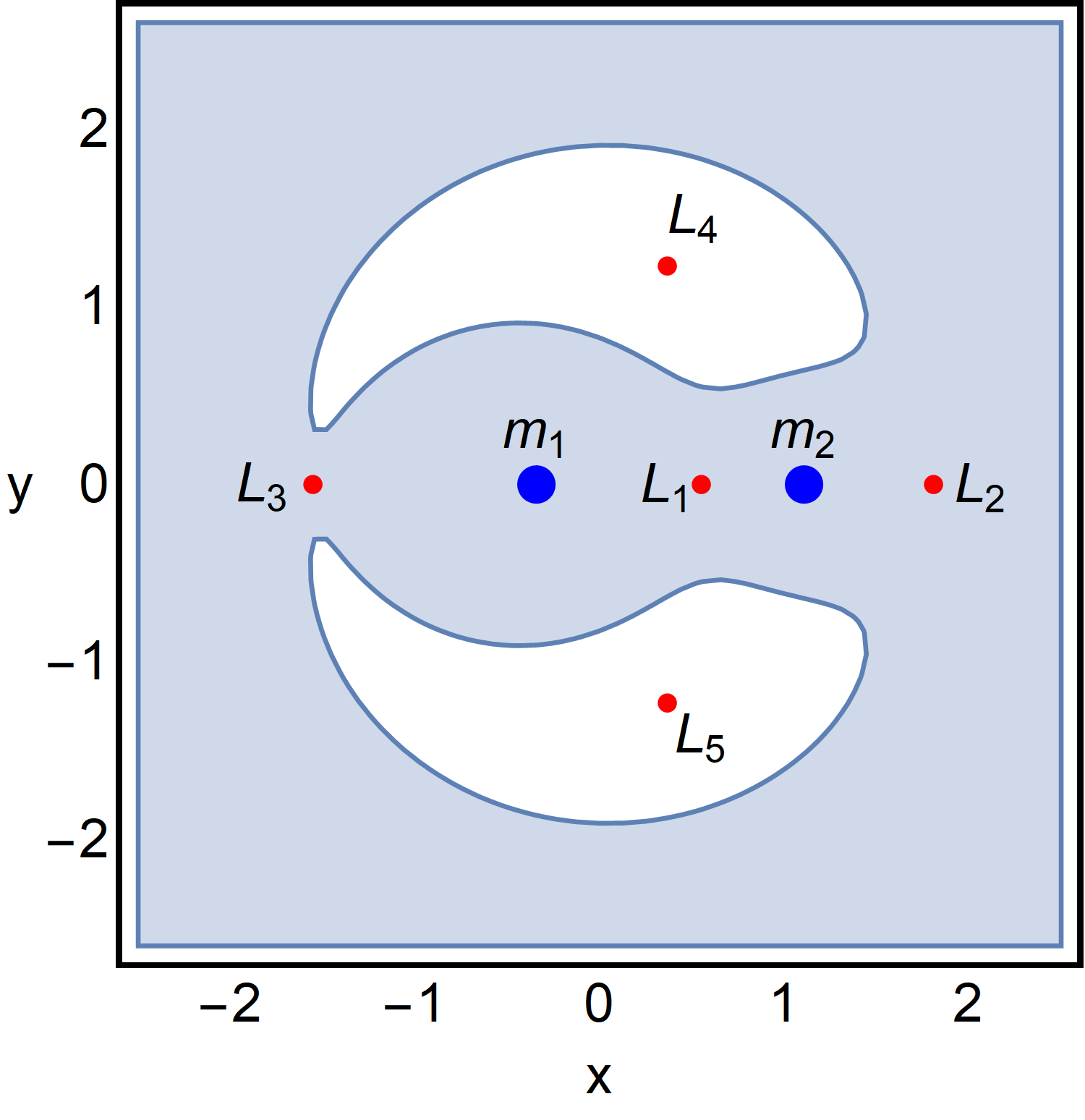}&
		\includegraphics[scale=.4]{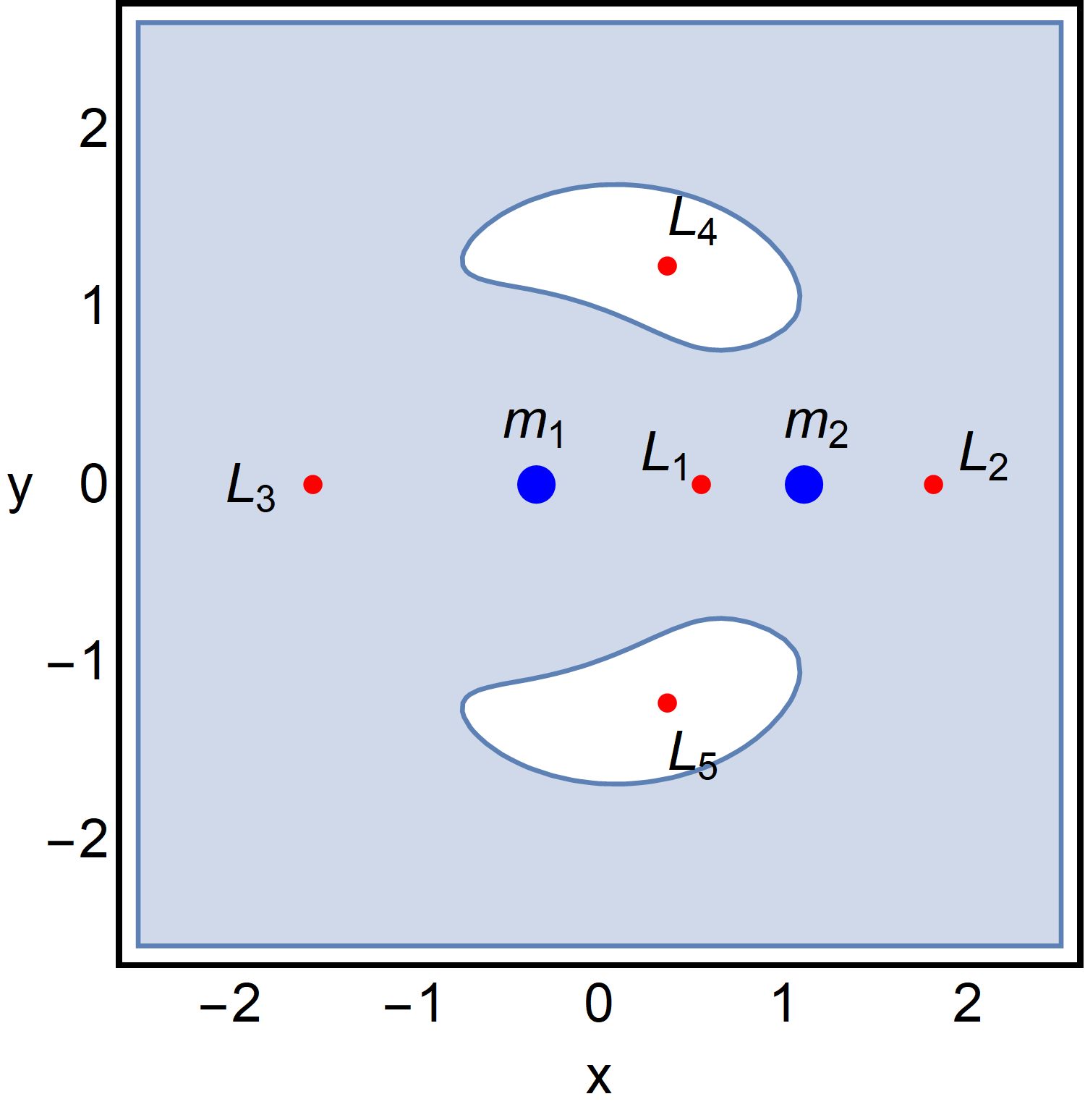}&
		\includegraphics[scale=.42]{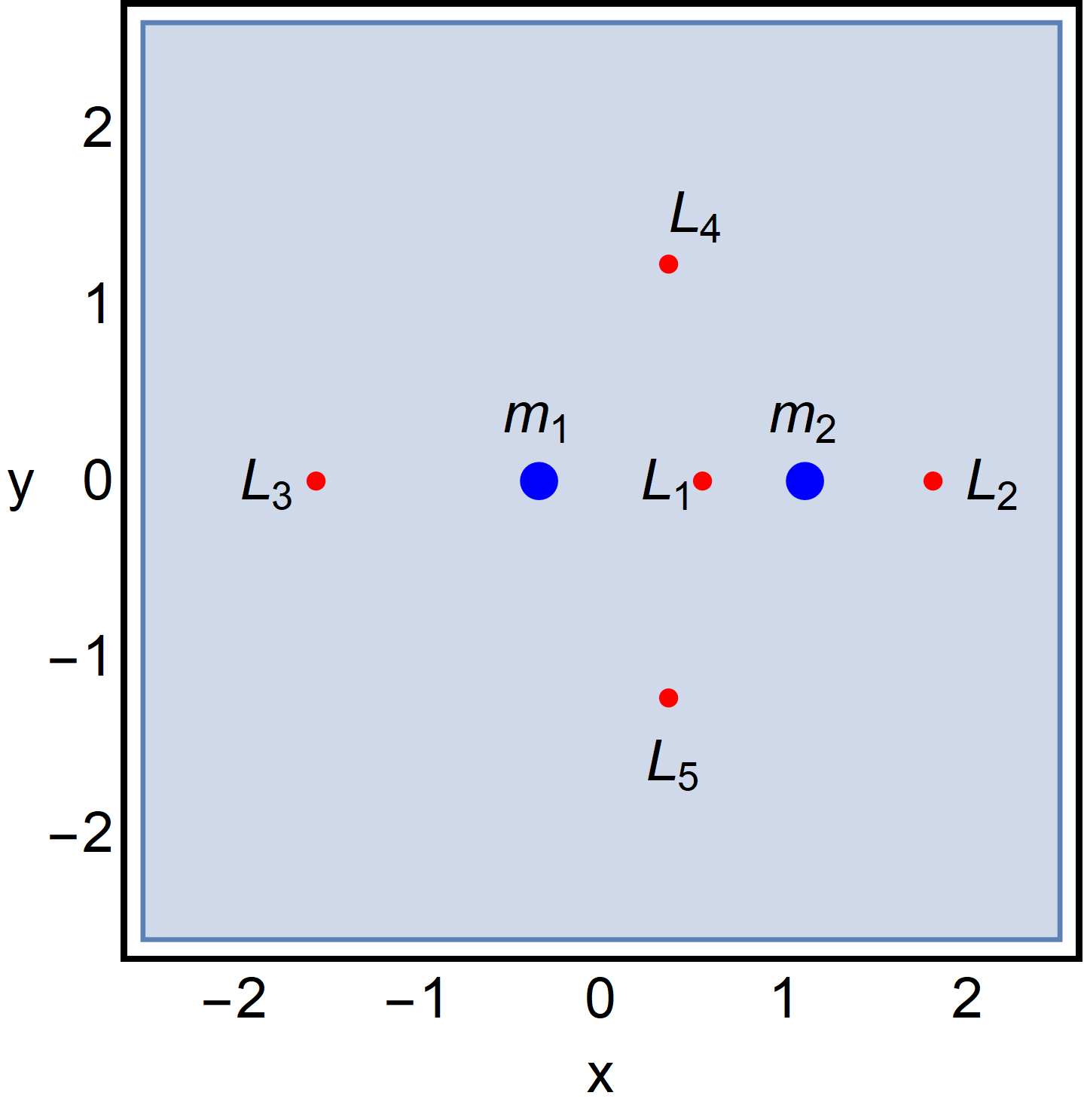} \\
		(d) C=5.95 &(e) C=5.55 &(f) C=5.15
	\end{tabular}
	\caption{Impact of variation of energy constant (Jacobi's Integral) on zero$-$velocity curves. Gray colour indicates region of motion. Blue color denotes the position of primaries $m_{1}$ and $m_{2}$. The position of libration points are denoted by black dots. Other parameters have values $M_{n1} = 0.08, M_{d1} = 2.0, M_{n2} = 0.04, M_{d2} = 0.6, a1 = 0.15, b1 = 0.2542,
		h1 = 0.00925, a2 = 0.175, b2 = 0.0789, h2 = 0.00875, cn1 = 0.0125, cn2 = 0.01, R = 1.45$.}
\end{figure*}
In the equation (9), if we take $\dot{x}=0$ and $\dot{y}=0$ then the expression $2 \Omega - C=0$ represents the zero-velocity curves. If we consider the expression $2 \Omega - C \geq 0$ then it will represent the permissible region of motion where the movement of the charged particle take place. 

In Fig. 4(a-f), the values of the fixed parameters are given. At these values, we find the position of all libration points. Five libration points ($\text{Li}$, i=1....5) are located. There are four different values of Ci's corresponding to Li's (Jacobi integral) as C4 and C5 are equal. The values of Ci's are $\text{C1}=6.86825,\text{C2}=6.5988, \text{C3}=5.96756$ and $\text{C4}=\text{C5}=5.22324$. We can divide the whole configuration space $(x,y)$ into three parts: interior regions, outlying regions and forbidden regions of motion. There are five different cases:
\begin{itemize}
	\item If $C>C1$ then all necks are closed and therefore orbit can move around primaries or in outlying regions. 
	\item When $ C2 < C \leq C1$, no necks are open but interior regions around $m_{1}$ and $m_{2}$ get connected. Therefore motion is possible around both primaries, i.e., around interior regions or in exterior regions. However, the orbit can not escape from inner  region to outer region or vice versa.
	\item When $C3 < C \leq C2$, one neck is open around $m_{2}$ allowing orbits to move close to  $m_{2}$ and escape to exterior region. 
	\item When $C4 < C \leq C3$, both necks are open; therefore, the movement of orbit can take place close to both primaries and escape to the outer region.
	\item When $C\leq C4$, the orbit can move through the whole configuration space.     
\end{itemize}
We have shown five different ((d) and (e) are of same cases) cases of zero velocity curves in Fig 2(a-f). 
In Fig. 2(a), we notice interior regions around $m_{1}$ and $ m_{2}$ and also the exterior regions. The orbit can move near primaries and not around any libration points. In Fig. 2(b), we see that the interior regions around $m_{1}$ and $m_{2}$ get connected. So the orbit can move in the locality of primaries $m_{1}$ and $m_{2}$. The orbit can also move around $L_{1}$. In Fig. 2(c), there is one exit channel around $m_{2}$ and $\text{L2}$. Therefore the orbit can move close to $m_{1}$ and $m_{2}$, and through exit channel, it can escape to outlying exterior regions. The movement of orbit is possible around $\text{L1}$ and  $\text{L2}$.  Now, we decrease the value of C (Jacobi's integral) gradually. In Fig. 2(d), at the value of $C=3.2$, there are two exit channels and also there are no forbidden regions around $m_{1}$, $m_{2}$, $\text{L1}$, $\text{L2}$ and $\text{L3}$. Thus we notice the increase in the region of motion, but still, the motion is not possible around the triangular liberation points. The same situation can be further observed in Fig. 2 (e). At the value $C=5.15$, the forbidden region of motion vanishes completely. Therefore the orbit can move anywhere in the configuration space $(x, y)$(see Fig. 2 (f)). Thus, we observe a significant impact on the geometry of zero-velocity curves due to variation of Jacobi's constant.

\section{Newton-Raphson(NR)-BoA  and Basin Entropy}
\subsection{NR-BoA}
We can determine different aspects of the dynamical system with the help of the NR-BoA. In the recent past, just a few researchers have applied this method in various dynamical system including different disturbing terms in the effective potential (for e.g.  \cite{Zotos2018}, \cite{Suraj2019}, \cite{Zotos17}, \cite{kalvouridis2012}, \cite{sprott2015} and their references). We use NR method (multivariate form) to study the BoA related to libration points. To reveal the domain of convergence for a particular libration point, we examine a set of initial conditions. To solve the systems of bivariate function $f(\textbf{X})=0$, we apply the iterative scheme
\begin{align*}
{\textbf{X}}_{n+1} ={\textbf{X}}_n-J^{-1}f(\textbf{X}_n),
\end{align*} where $J$ is the Jacobian matrix of $f(\textbf{X}_n)$.\\
In this work, the system of differential equations are given by
\begin{align*}
\Omega_{x} = 0,\\
\Omega_{y} = 0.
\end{align*}
With elementary calculations, we get the iterative formula for each coordinate as:
\begin{eqnarray}\nonumber
{x}_{n+1}={x}_n-\left(\frac{\Omega_{x_n} \Omega_{{y_n}{y_n}}-\Omega_{y_n} \Omega_{x_ny_n}}{\Omega_{x_nx_n}\Omega_{y_ny_n}-\Omega_{x_ny_n}\Omega_{y_nx_n}}\right), \quad
{y}_{n+1}={y}_n+\left(\frac{\Omega_{x_n} \Omega_{{y_nx_n}}-\Omega_{y_n} \Omega_{x_nx_n}}{\Omega_{x_nx_n}\Omega_{y_ny_n}-\Omega_{x_ny_n}\Omega_{y_nx_n}}\right),
\end{eqnarray}
where ${x}_n$,  and ${y}_n$ are $n$-th step of the NR method. Partial derivatives of $\Omega(x,y)$ are given in the form of subscripts.
The algorithm for computing BoA, we have applied the following steps:
\begin{itemlist}
	\item Initial conditions $({x}_0,{y}_0)$ are taken on the configuration plane $(x, y)$ and apply NR method. In present calculations, we have adopted a grid of $1024\times1024$ initial conditions in an uniform way. These initial conditions are called nodes. The Min. and Max. values of ${x}$ and ${y}$ are chosen to view the complete picture of the BoA generated by the libration points.
	\item The method is applied continuously till an accuracy of order $10^{-15}$ or the maximum number of iterations (500) is reached for each initial condition. The stopping condition $|\textbf{x}_{n+1}-\textbf{x}_{n}|\leq 10^{-15}$ or $N\leq 500$ is used. 
	\item For each initial condition, we record the number of iterations $\text{N}$ to achieve the desired accuracy. For non-converging initial conditions, we further iterate for 10000 iterations using this method. 
	\item We fix different colours for each libration point, and a particular colour is assigned to the initial conditions according to its convergence towards a specific libration point.
	\item  After assigning all initial conditions one colour, we plot the colour coded graph, which is known as BoA. For all computations and simulations, we have used Mathematica 11.0 \cite{wolf2014}.
\end{itemlist}       
\subsection{Basin Entropy}
In 2016, A. Daza \cite{Alvar16} introduced a new tool to measure unpredictability of the basin of attraction. This new tool can quantify the uncertainty of BoA, known as basin entropy. We shall briefly discuss the algorithm for the computation of basin entropy:
\begin{itemlist}    
	\item First of all, we complete the process of plotting BoA. In the configuration plane $(x, y)$,  1024$\times$1024 initial conditions are taken, each having some colour as per its convergence towards libration points.
	
	\item In this step, we divide the whole region into different non-overlapping boxes to completely cover up the entire area. Each box contains precisely 25 trajectories. We have considered approximately 42000 non-overlapping boxes for this computation.
	
	\item We compute the probability of colour $j$ inside each box $i$ denoted as $p_{ij}$. The gibbs entropy for each box $i$ is computed as 
		\begin{equation}
		\text{S}_{i}= \sum_{j=1}^{m_{i}} \text{p}_{ij} \log\left(\frac{1}{\text{p}_{ij}}\right), 
		\end{equation}
		where $m_{i}\in [1,\text{N}_{A}]$ is the number of colours inside the box $i$ and $\text{N}_{A}$ represents the number of libration points. The probabilities $\text{p}_{ij}$ are calculated as 
		\begin{equation}\nonumber
		\text{p}_{ij}=\frac{\text{number of trajectories leading to colour j} }{\text{number of trajectories in the box i}}.
		\end{equation}
		\item Eventually, due to selection of non overlapping boxes $\text{N}$, the entropy of the whole grid is equal to the summation of entropy of each box $i$ of the grid.
		\begin{equation*}
		\text{S}=\sum_{i=1}^{\text{N}} \text{S}_{i}= \sum_{i=1}^{\text{N}} \sum_{j=1}^{m_{i}}\text{p}_{ij} \log\left(\frac{1}{\text{p}_{ij}}\right).
		\end{equation*}
		Now, we define the basin entropy as
		\begin{equation*}
		\text{S}_{b}=\frac{\text{S}}{\text{N}}.
		\end{equation*}
		In similar way, we define boundary basin entropy as
		\begin{equation*}
		\text{S}_{bb}=\frac{\text{S}}{\text{N}_{b}}.
		\end{equation*}
		where $\text{N}_{b}$ denotes the number of boxes having more than one colour.  
	\end{itemlist}
	If the values of $\text{S}_{b}$ and $\text{S}_{bb}$ are greater than $\log{2}$, the BoA or boundaries along BoA is fractal in nature. 
	Now, we study the effect of the parameters $\text{R}, \text{M}_{n1}, \text{M}_{d1}, \text{M}_{n2} \ \text{and} \ \text{M}_{d2}$ on the BoA. We have considered two cases: In first case, the effect of  parameter $R$ and in the second case, effect of parameters $\text{M}_{n1}, \text{M}_{d1}, \text{M}_{n2}\  \text{and}  \ \text{M}_{d2}$. 
	\section{Effect of parameters $\text{R}\ \& \ \text{M}_{n1}, \text{M}_{d1}, \text{M}_{n2} \ \text{and}$  $\text{M}_{d2}$  on the BoA}
	\subsection{Influence of $R$ (distance between center of two galaxies)}
	
	\begin{figure*}[htb!]
		\centering
		\begin{tabular}{cc}
			\includegraphics[scale=.4]{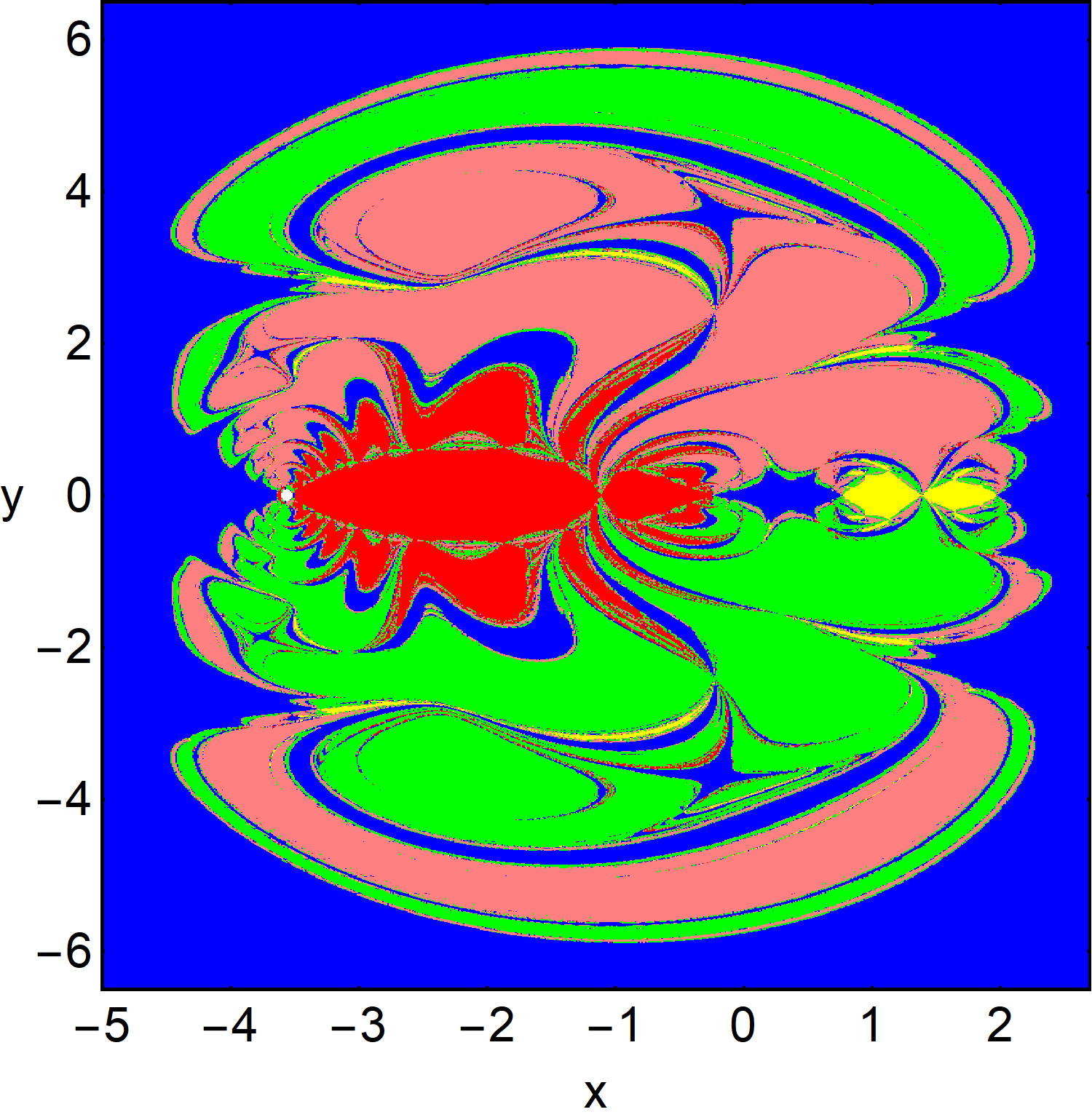} ~~~~~~~~~~~&
			\includegraphics[scale=.4]{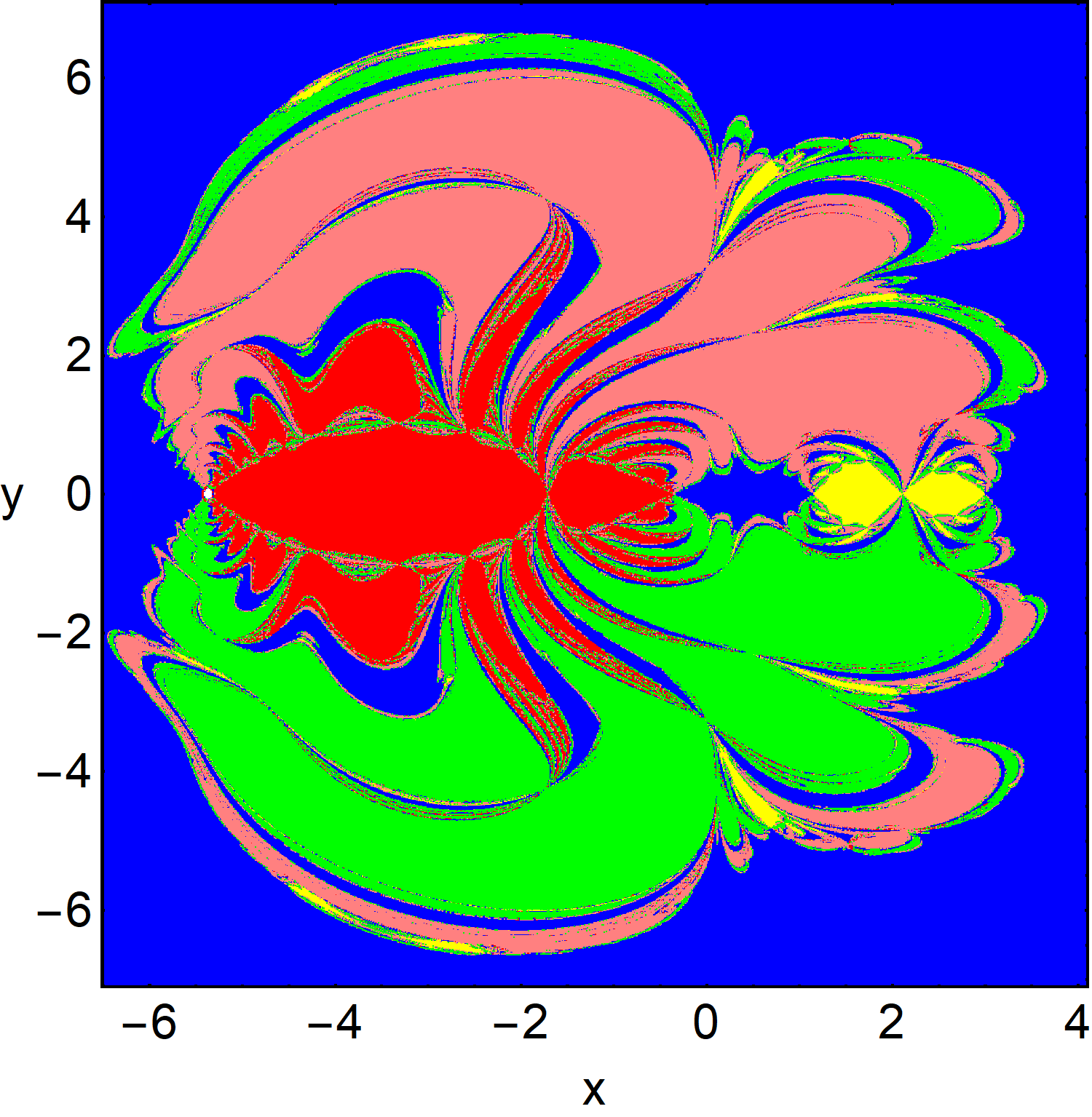}\\
			(a) $R$=1~~~~~~~~~ &(b) $R$=1.5 \\
			{}&{}\\
			\includegraphics[scale=.4]{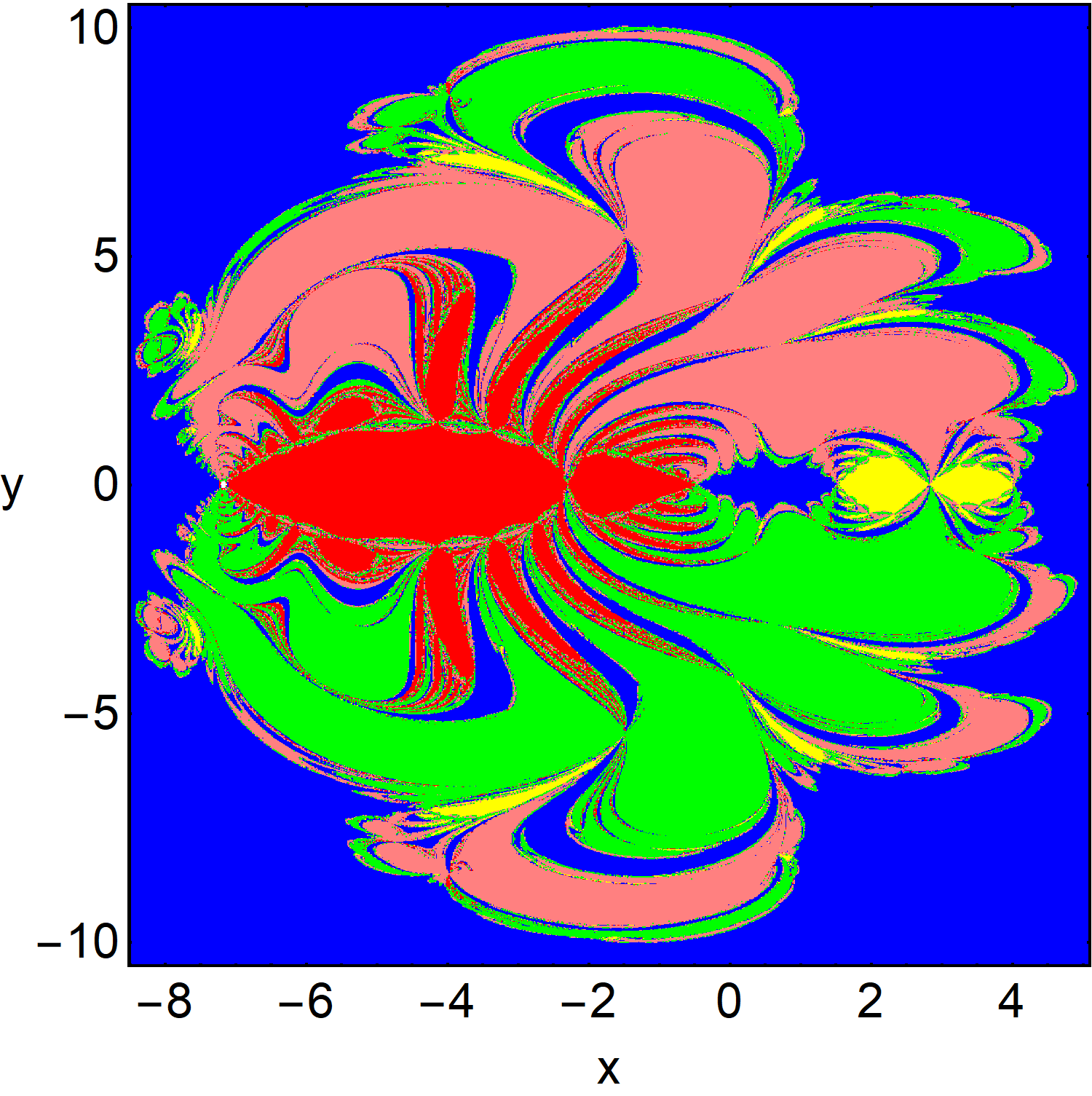}~~~~~~~~~~~&
			\includegraphics[scale=.4]{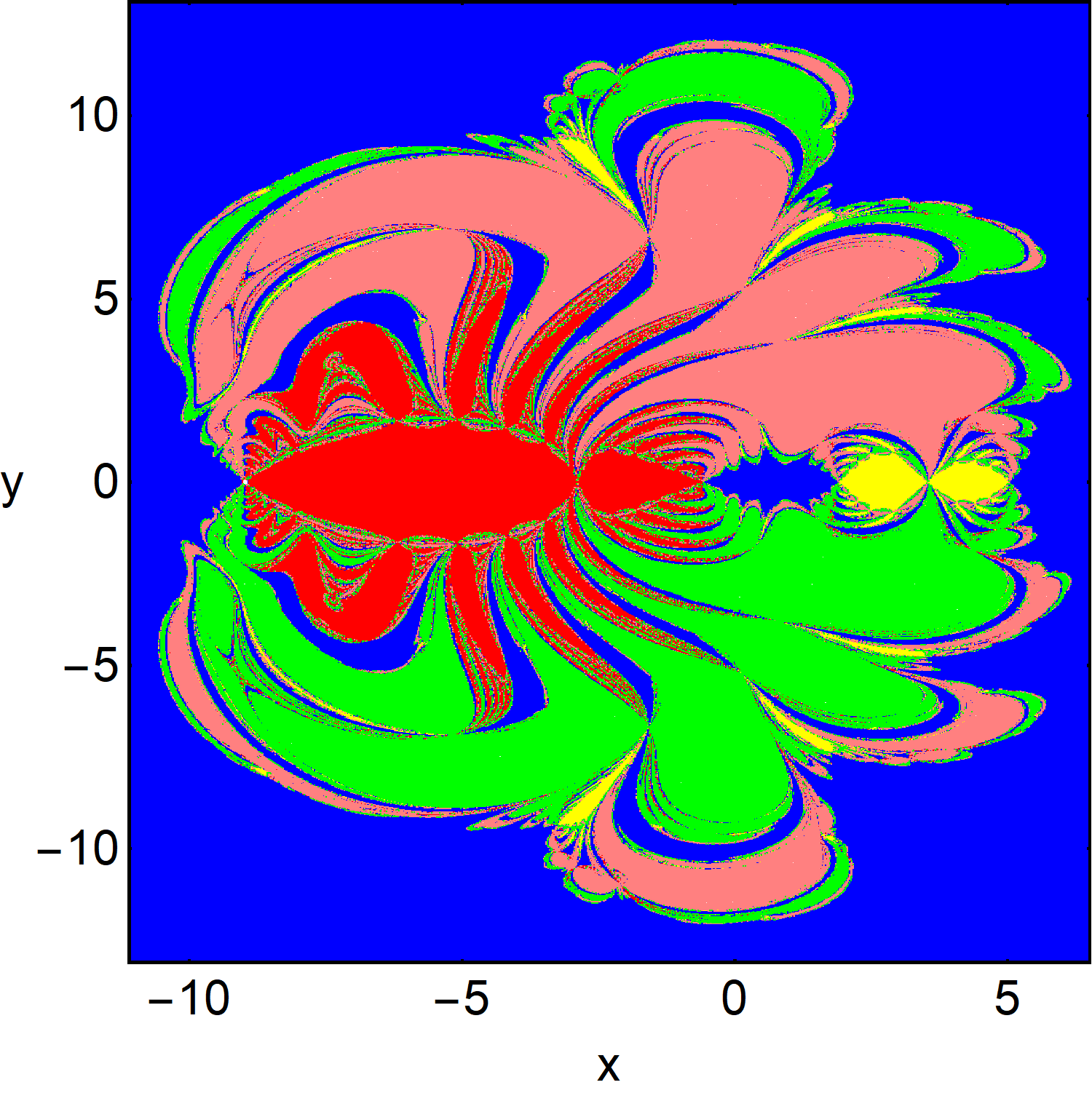}\\		
			(c) $R$=2~~~~~~~~~&(d) $R$=2.5\\
			{}&{}\\
			\includegraphics[scale=.4]{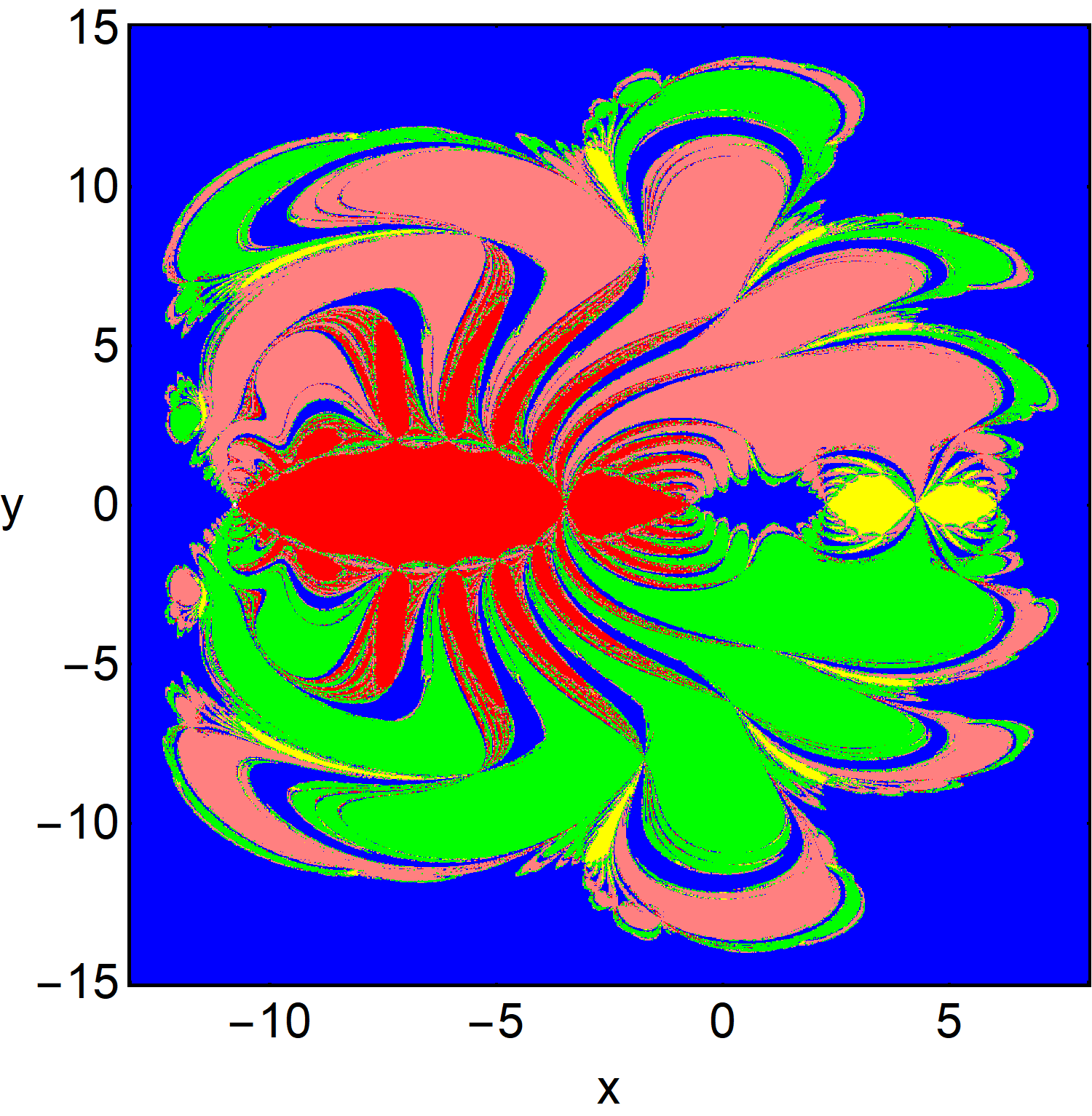}~~~~~~~~~~~~&
			\includegraphics[scale=.4]{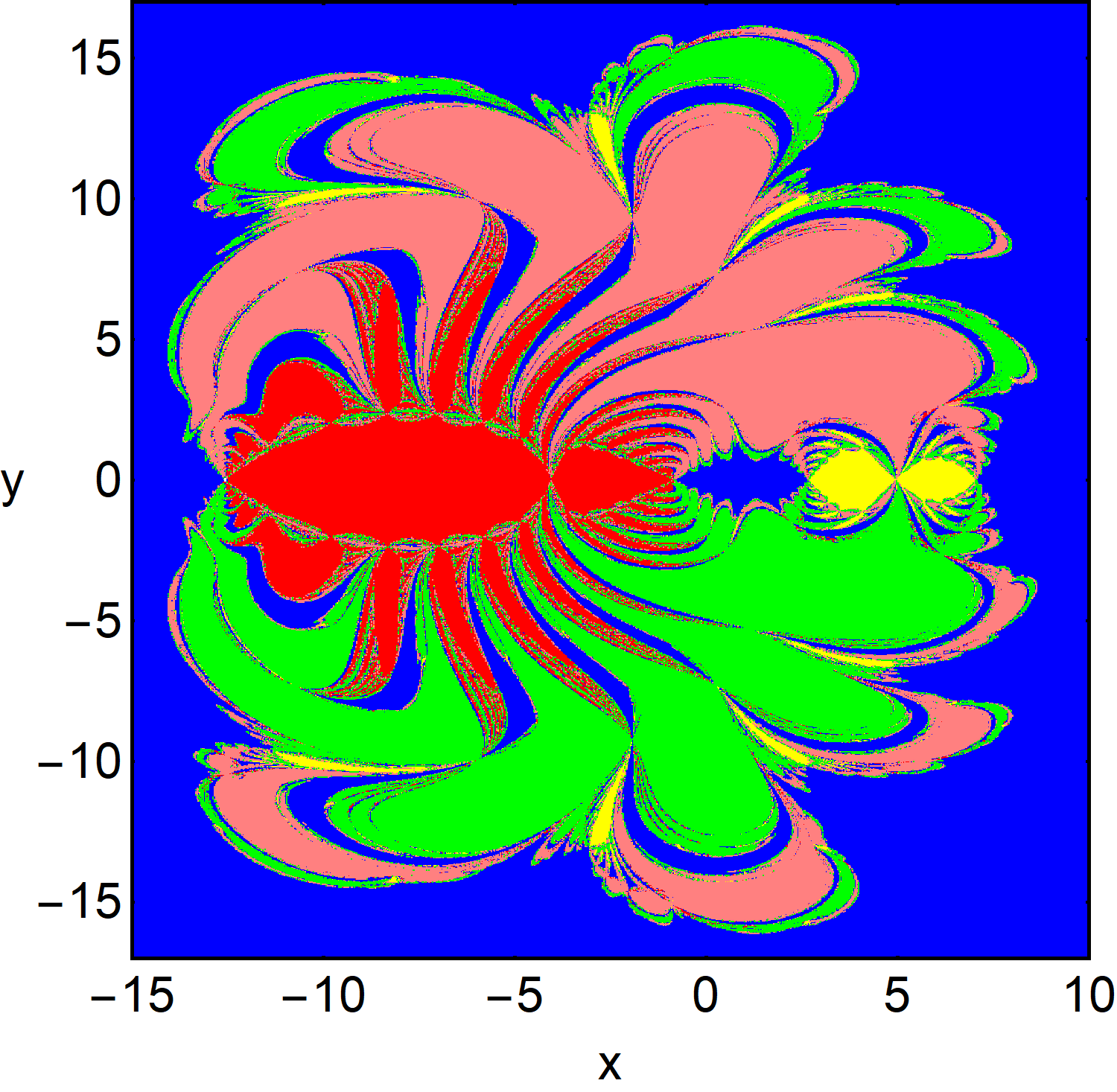}\\		
			(e) $R$=3~~~~~~~~~~&(f) $R$=3.5\\
		\end{tabular}
		\caption{BoA for different values of $R$. The colour codes for the BoA related to libration points are $\text{L}1$ (blue), $\text{L}2$ (Red), $\text{L}3$ (Yellow) , $\text{L}4$ (Pink) and $\text{L}5$ (Green).}
	\end{figure*}

\begin{figure*}[htb!]
	\centering
	\begin{tabular}{cc}
		\includegraphics[scale=.57]{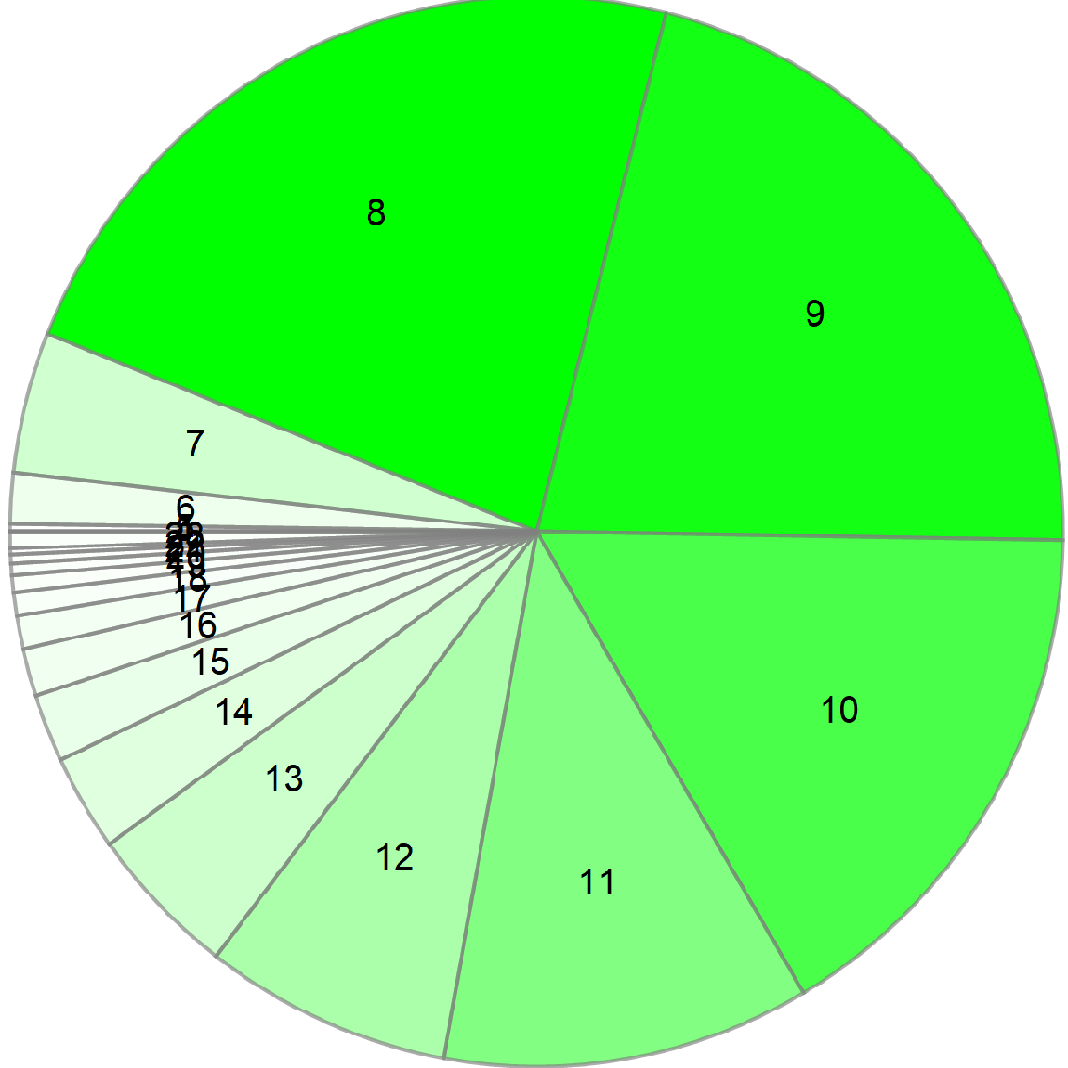}~~~~~~~~~&
		\includegraphics[scale=.5]{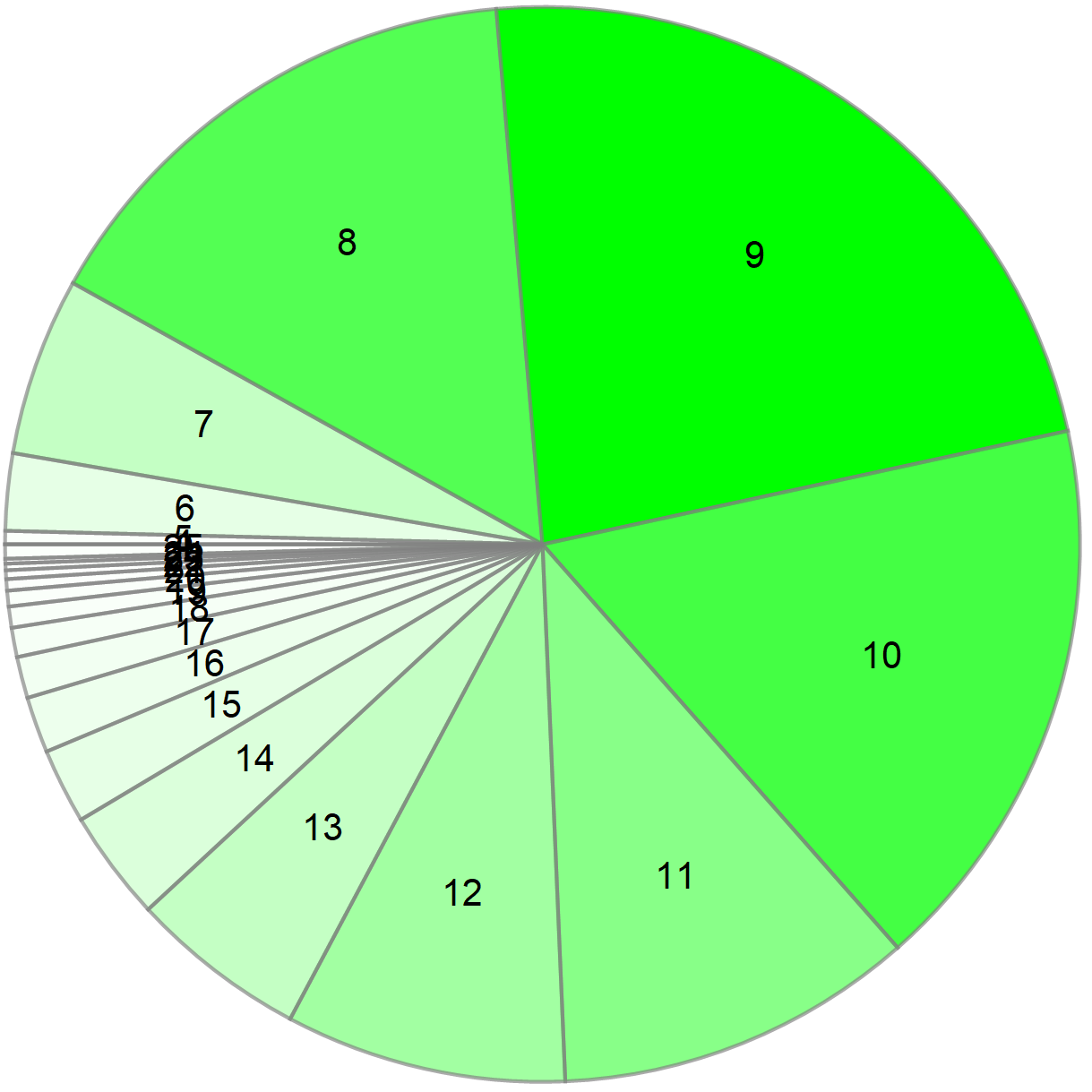}\\
		(a) $R$=1, MINo.=8 ~~~~~~~~~&(b) $R$=1.5, MINo.=9 \\
		{}&{}\\
		\includegraphics[scale=.5]{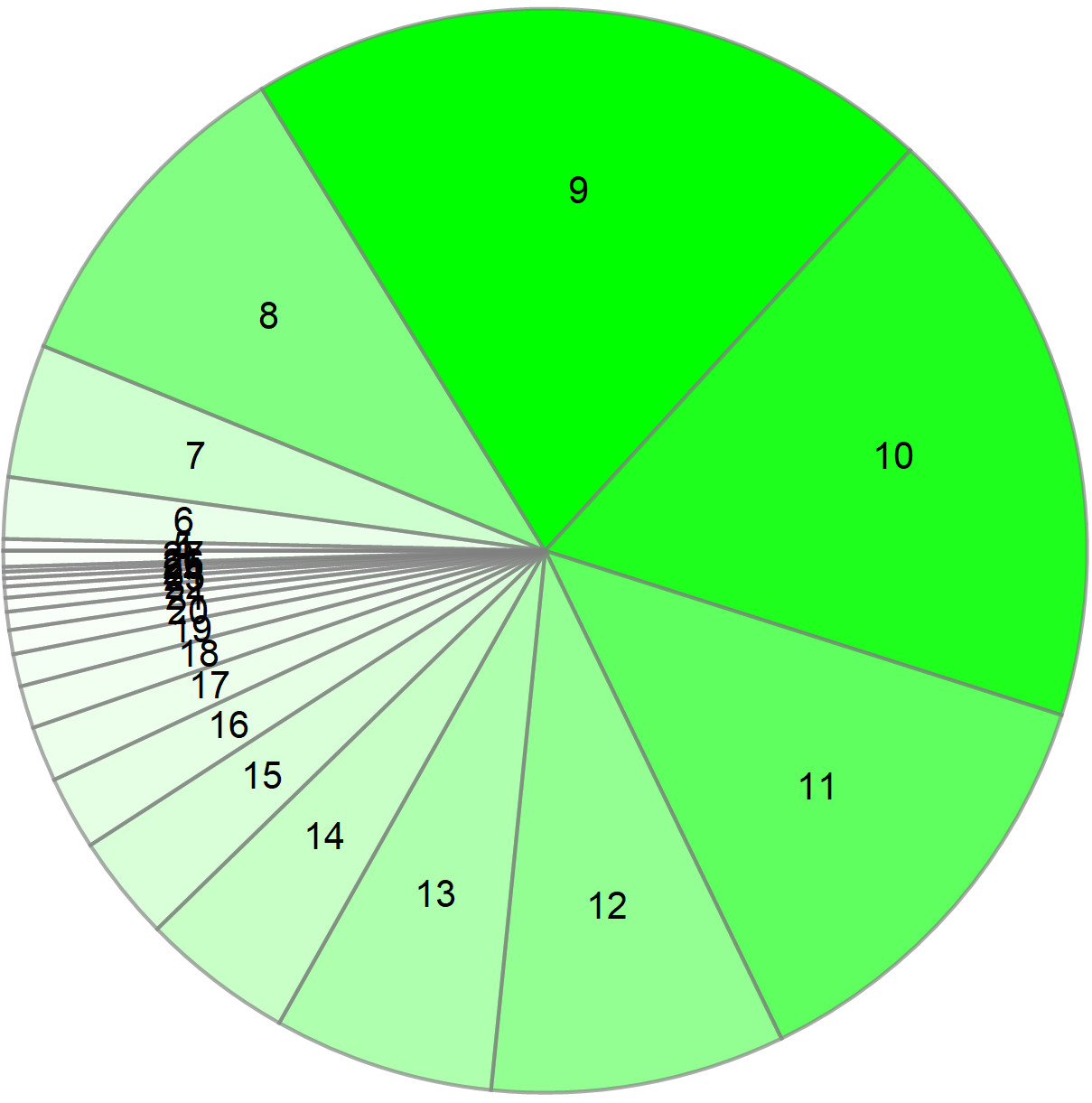}~~~~~~~~~~&
		\includegraphics[scale=.5]{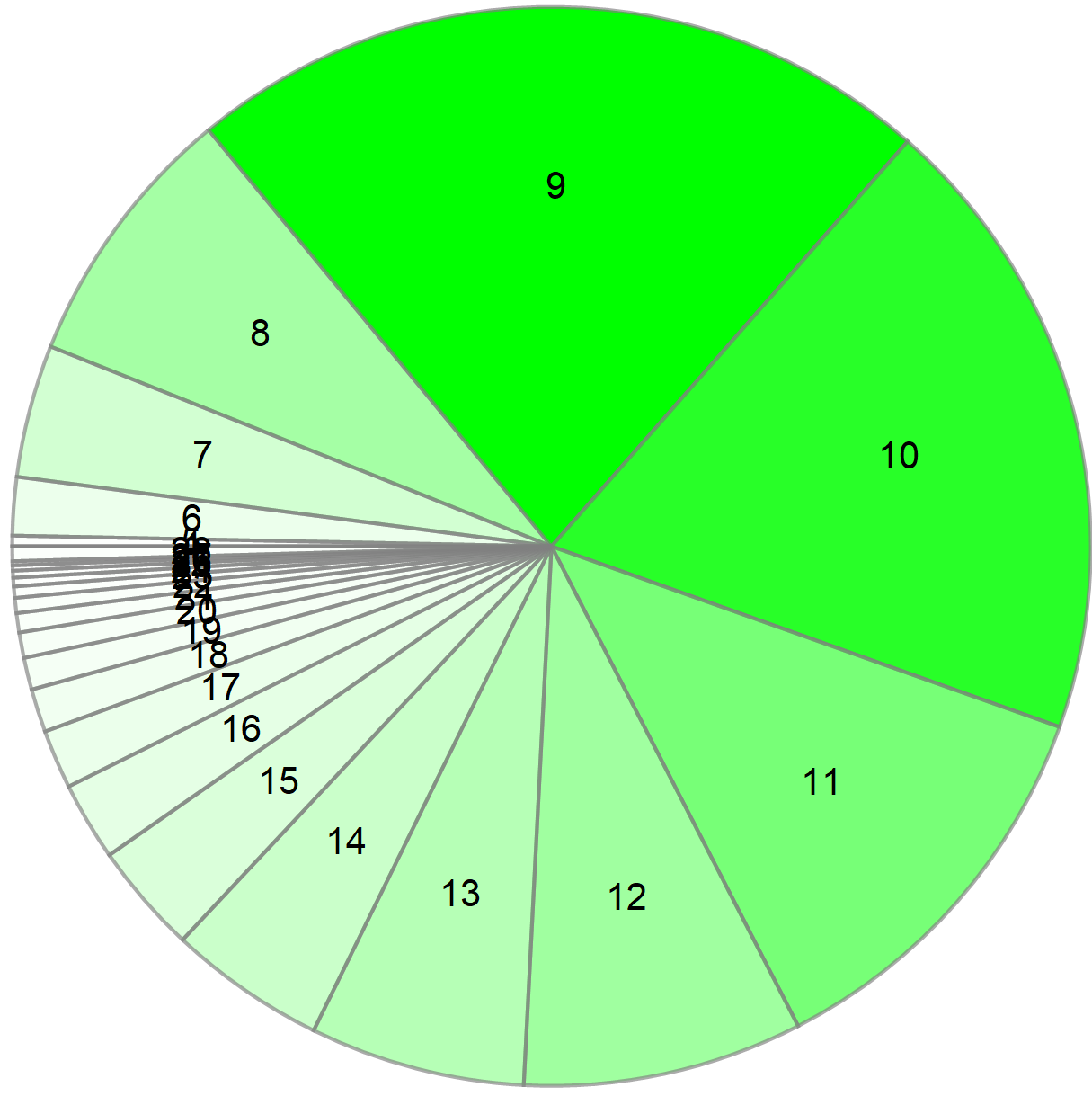}\\		
		(c) $R$=2, MINo.=9~~~~~~~~~~&(d) $R$=2.5, MINo.=9\\
		{}&{}\\
		\includegraphics[scale=.5]{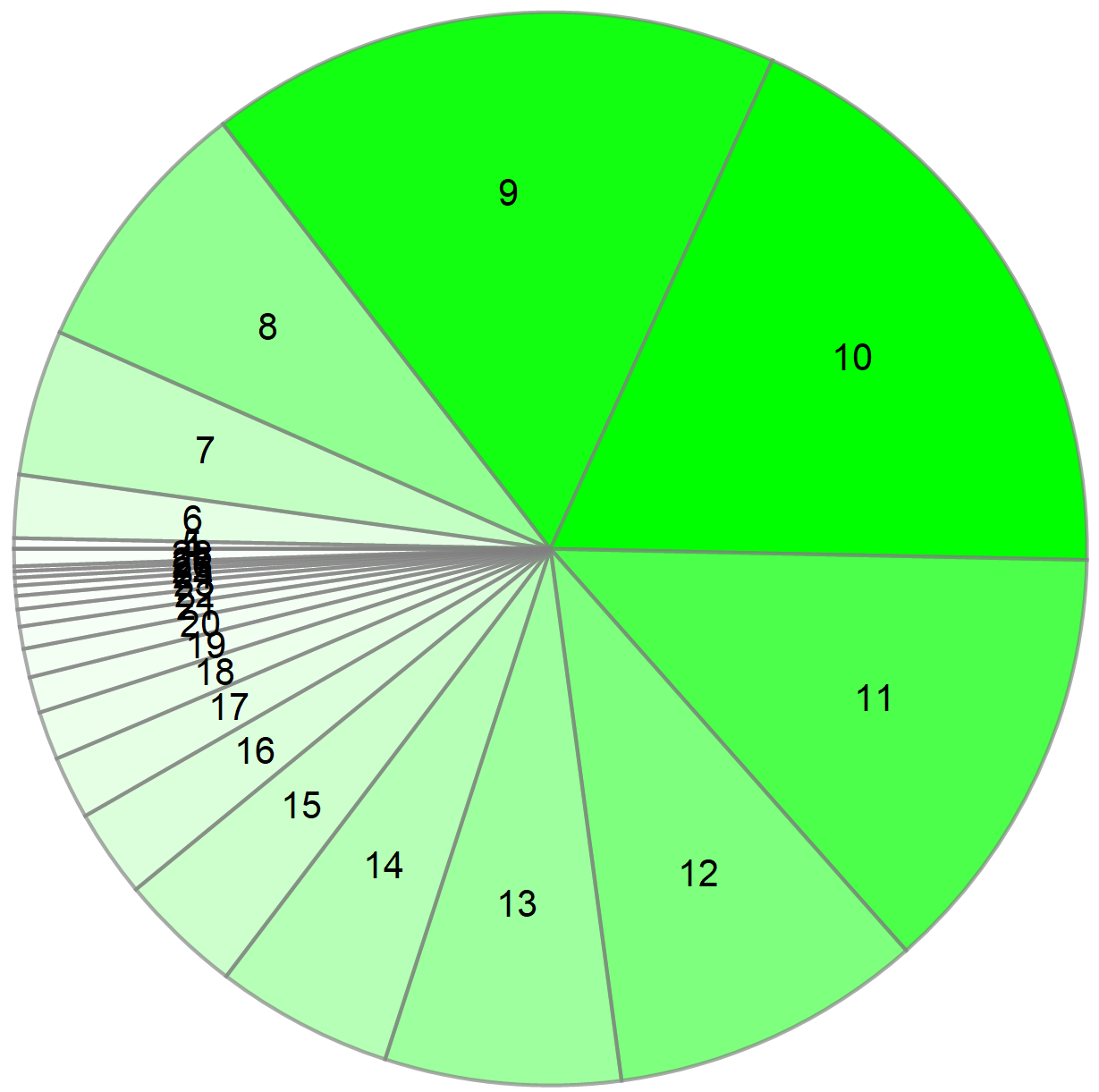}~~~~~~~~~~&
		\includegraphics[scale=.5]{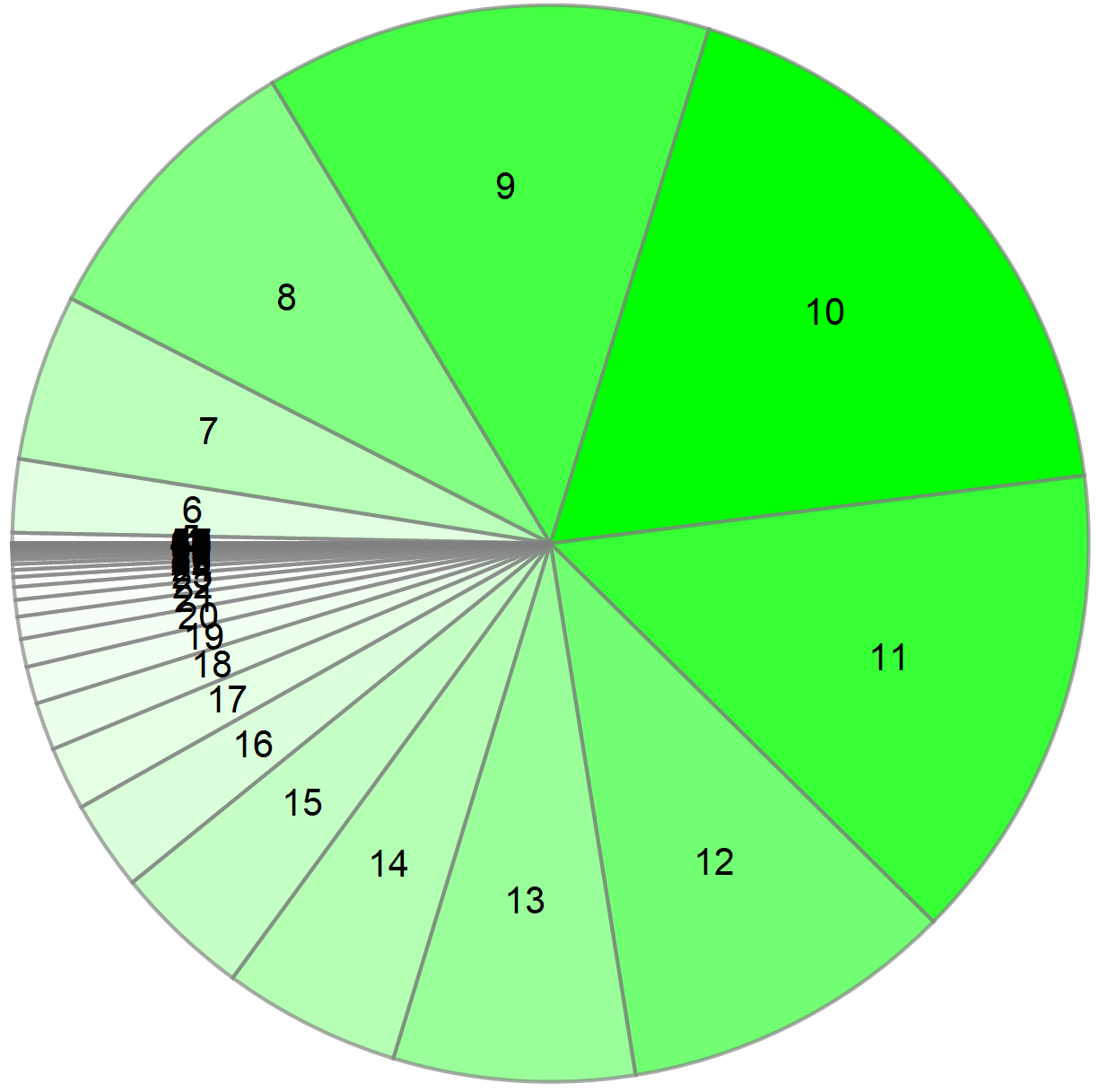}\\		
		(e) $R$=3, MINo.=10~~~~~~~~~~&(f) $R$=3.5, MINo.=10\\
	\end{tabular}
	\caption{(a-f) $\text{Pie-chart}$ of number of required iterations for different values of $R$. MINo. denotes the maximum number of  iterations.}
\end{figure*}

	\begin{figure*}[htb!]
		\centering
		\begin{tabular}{cc}
			\includegraphics[scale=.4]{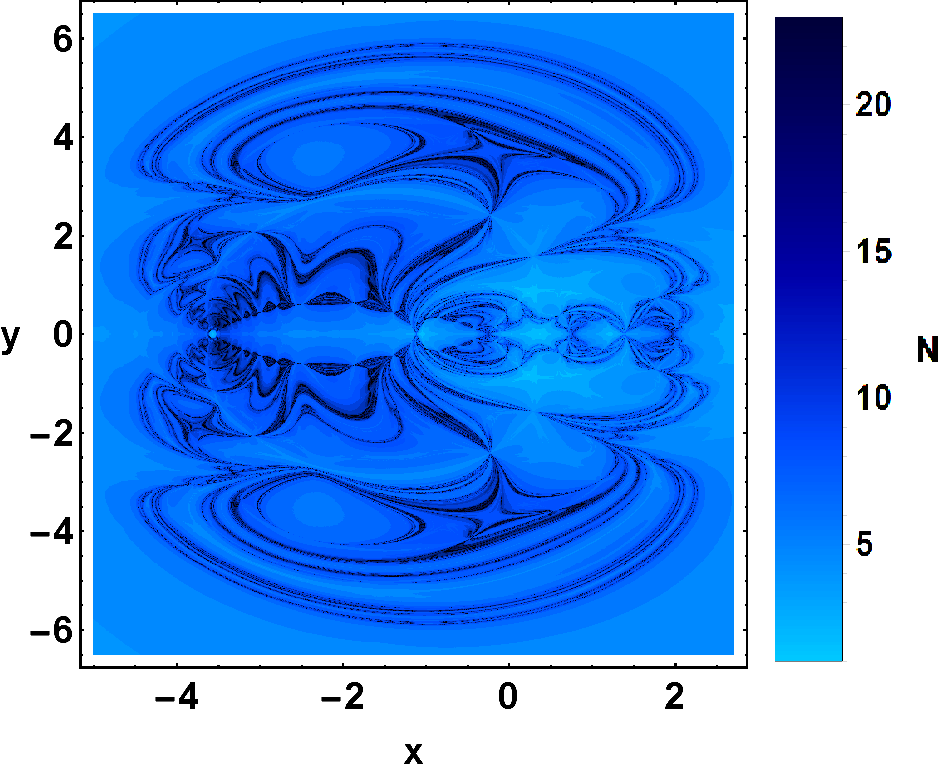}~~~~~~~ &
			\includegraphics[scale=.4]{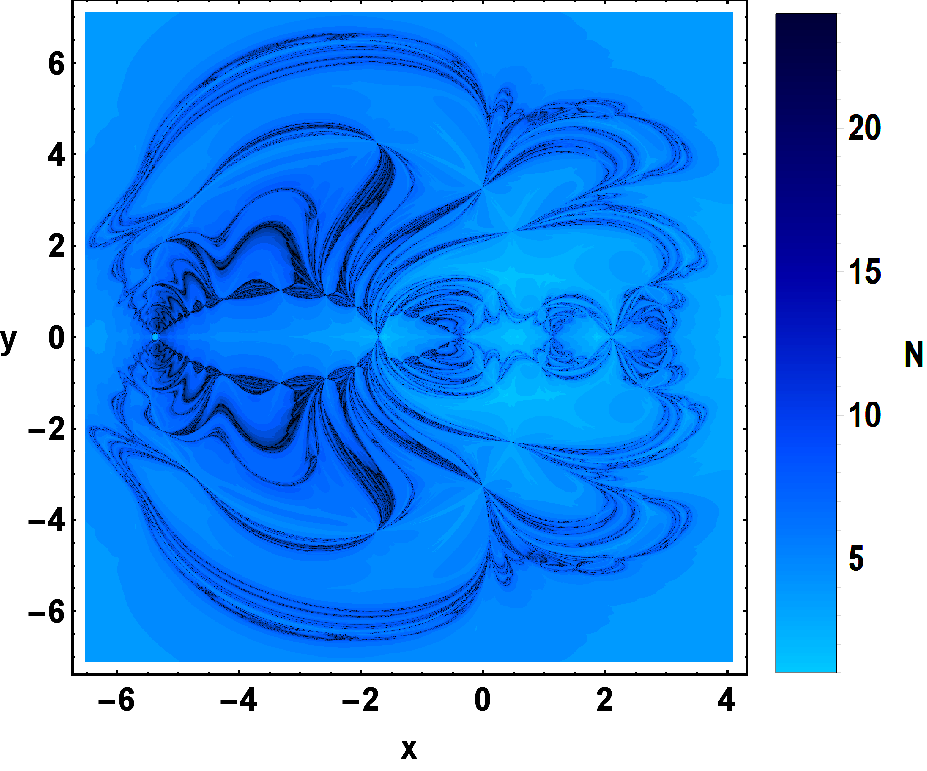}\\
			(a) $R$=1 &(b) $R$=1.5 \\
			{}&{}\\
			\includegraphics[scale=.26]{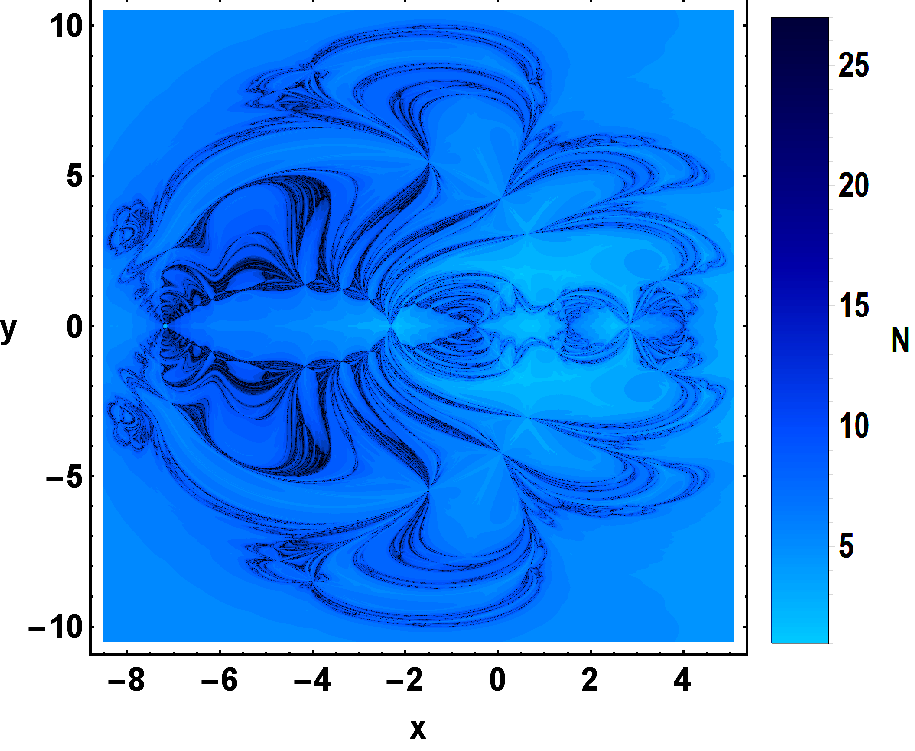}~~~~~~~&
			\includegraphics[scale=.4]{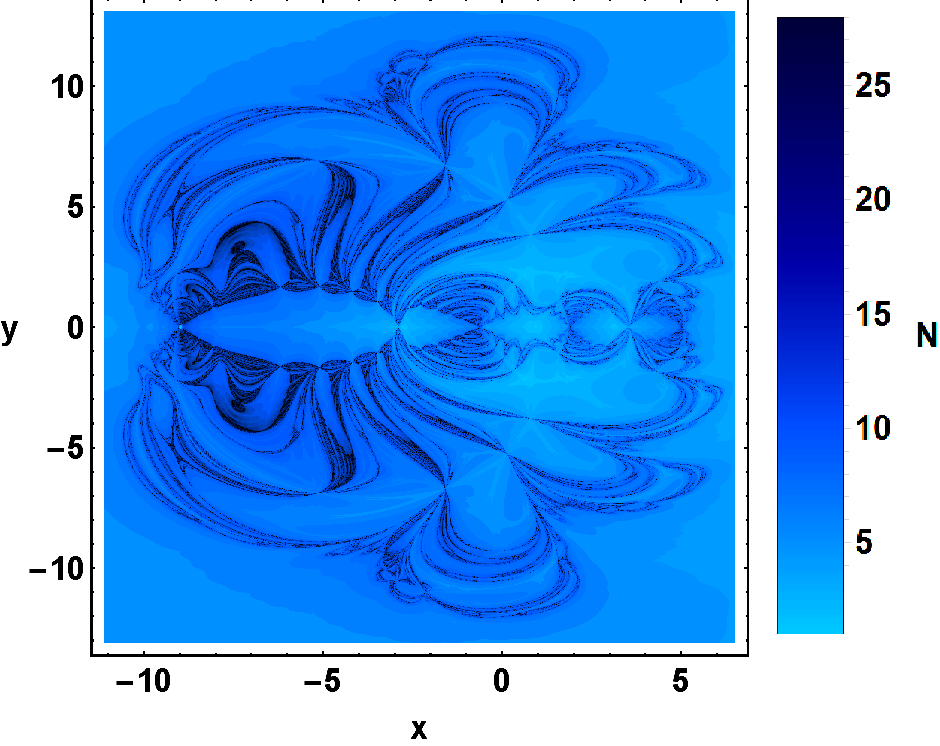}\\		
			(c) $R$=2 & (d) $R$=2.5\\
			{}&{}\\
			\includegraphics[scale=.4]{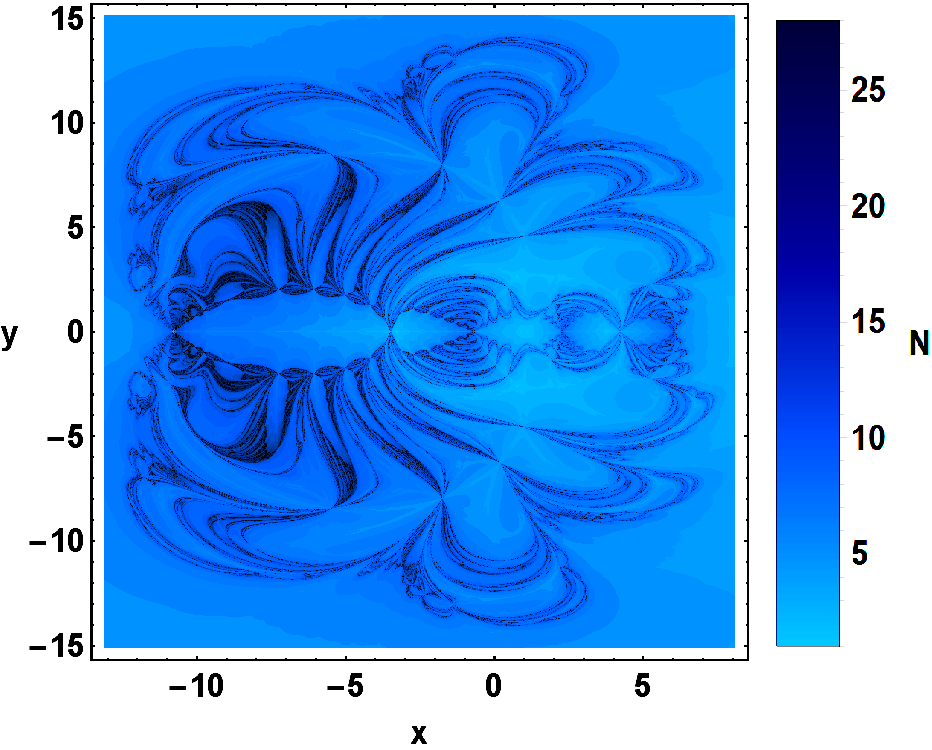}~~~~~~~&
			\includegraphics[scale=.4]{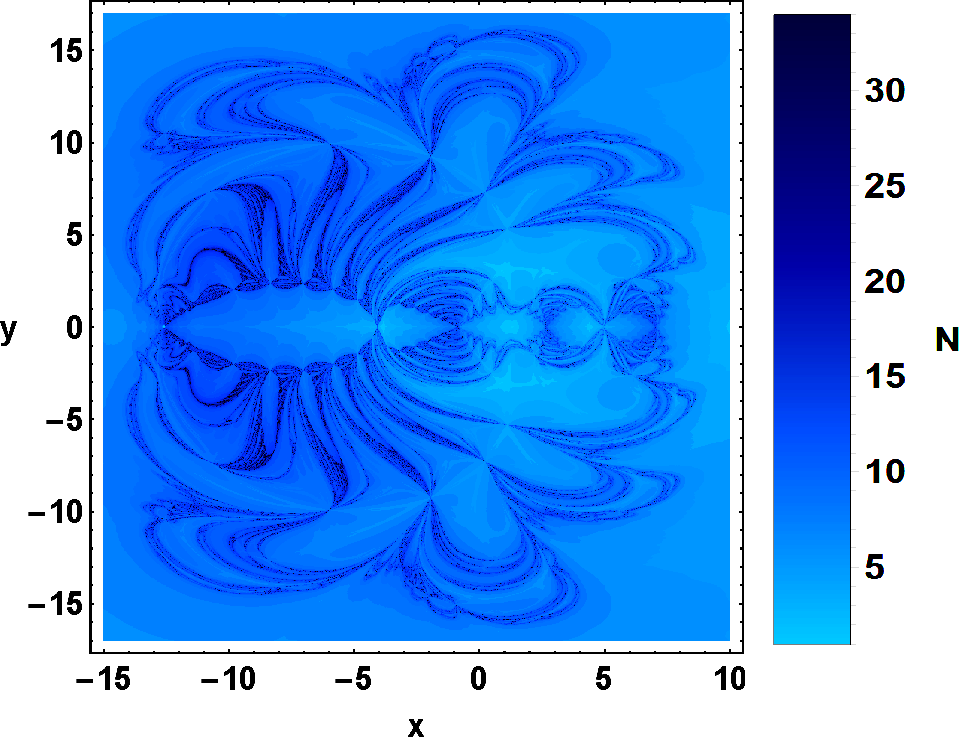}\\		
			(e) $R$=3&(f) $R$=3.5\\
		\end{tabular}
		\caption{(a-f) Iterations \textbf{N} (blue tone) required to achieve desired accuracy for different values of R.}
	\end{figure*}

	\begin{table}
		\tbl{Number of initial conditions converging towards different libration points and the values of  entropy $\text{S}_{b}$ and $\text{S}_{bb}$ due to variation in parameter $R$.}
		{\begin{tabular}{|c|c|c|c|c|c|c|c|c|c|}
				\hline
				$R$ & $\text{Total Points}$ & $\text{L1}$& $\text{L2}$ & $\text{L3}$& $\text{L4}$ &$\text{L5}$ & Execution time &$\text{S}_{b}$ & $\text{S}_{bb}$ \\ [2pt] 
				\hline 
				1.0 & 1003072  &  453456   & 10797   & 69495  & 234662  & 234662  & 2029.3  & 0.923136  & 1.07934 \\ [1pt]
				\hline
				1.5 &1047741  &  465262   & 107185 & 17873  & 228770  & 228651  & 2809.01& 0.884452  & 1.10743   \\ [1pt] 
				\hline
				2.0 &1051051  &  499101   & 97712   & 20272  & 216922  & 217044  & 3539.67& 0.806377  & 0.930766 \\ [1pt]
				\hline
				2.5 &1046999  &  522472   & 104492 & 18384  & 200762  & 200889  & 3161.13& 0.881633  & 0.991321 \\ [1pt] 
				\hline
				3.0 &1024382  &  486481   & 98751   & 20294  & 209429  & 209427  & 2149.5  & 0.919124  & 1.0552     \\ [1pt] 
				\hline
				3.5 &1012300  &  476105   & 104206 & 19413  & 206381  & 206195  & 3410.12& 0.940689  & 1.07509   \\ [1pt]	
				\hline
		\end{tabular}}   
		
	\end{table}

    We begin our investigations with the case, where $R$ varies in the interval $[1, 3.5]$ and we take $\text{M}_{n1} = 0.08, \text{M}_{d1} = 2.0,\text{M}_{n2} = 0.04$ and $\text{M}_{d2} = 0.6$. In Fig. 3, the BoA for libration points are depicted by five different colours. Also, the BoA cover all of the configuration plane $(x, y)$. Most importantly, we notice that a slight change in the value of the parameter $R$, the geometry of the basins changes significantly. In particular, the change in Fig. 3 (a) to Fig. 3(b) can be seen. Here the region of liberation point $\text{L5}$ (green) in the upper part of Fig. 3 (a) shrunk significantly in Fig. 3 (b) due to little change in the value of $R$. A similar observation can be seen for the basin corresponding to liberation point $\text{L4}$ (pink) in the bottom of the figures. When the value of $R$ is changing from $1$ to till $3.5$, a major change is observed in Fig. 3(a) to Fig. 3 (f). 
   Despite having the same number of initial conditions converging towards $\text{L4}$ and $\text{L5}$, there is a regular change in the shape of BoA for different values of $R$. The domain of convergence of libration point $\text{L1}$ extends towards infinity whereas the domain of convergence due to libration points $\text{L2, L3, L4}$ and $\text{L5}$ is finite for such values of $R$. Further, we observe that basins for libration points $\text{L2}$ and $\text{L3}$ have the shape like insect with legs and antennas. Also, the shape of BoA corresponding to libration points $\text{L4}$ and $\text{L5}$ appears as many wings of a butterfly. 
   In Fig. 4 (a-f), we have shown Pie-Chart for the number of  iterations needed for all initial conditions taken in the configuration plane $(x, y)$. In Fig. 4 (a), it is clear that the maximum number of  iterations needed are 8, 9 and 10 $(8\to 229994, 9\to 213628, 10\to 163375)$. Likewise, we have recorded the number of iterations taken by initial conditions to converge for different values of $R$. We notice that approximately 95 \% of the initial conditions converge after 25 iterations. A maximum number of initial conditions converge after 8 to10 number of iterations. 
   
   In Fig. 5 (a-f) the relationship between numbers of iterations needed and the initial conditions considered in the configuration plane $(x, y)$ has been established. The blue tone is used to represent the iterations needed for each initial condition. We notice that the initial conditions along the boundaries of the basins require more iterations than other initial conditions. In Table 1, we have mentioned the details of convergence of initial conditions towards the libration points $\text{L1}$, $\text{L2}$, $\text{L3}$, $\text{L4}$  and $\text{L5}$. The value of basin entropy $\text{S}_{\text{b}}$ and boundary basin entropy $\text{S}_{\text{bb}}$ is also given in Table 1 for different values of $R$. 
   Based on Table 1, Fig. 3, Fig. 4 and Fig. 5, we sum up the following things:

   \begin{itemize}
   	\item Due to changes in the value of $R$, if we move from Fig. 3(a) to Fig. 3(f), we observe a significant change in the geometry of BoA.  
   	\item The region occupied by the initial conditions converging towards $\text{L4}$ and $\text{L5}$ are approximately equal for each case, although their geometry is changing in every case (see Table 1).
   	\item In all cases, the maximum number of iterations required for the convergence of $95 \%$ of the initial conditions are below 25. To achieve the desired accuracy, i.e. of order $10^{-15}$, the maximum probable number of iterations lies between 8 to 10 (See Fig. 4(a-f)). 
   	\item From Table 1, we see that the value of basin entropy  $\text{S}_{\text{b}}$ and boundary basin entropy  $\text{S}_{\text{bb}}$ is greater than $\log{(2)}$ and therefore the existence of fractal is verified. The value of boundary basin entropy  $\text{S}_{\text{bb}}$ is higher than the entropy of other regions indicates the existence of higher fractal regions along the boundaries. From Table 1 and Fig. 5, we conclude that the initial conditions which lie within the fractal regions require more iterations to converge.  
   \end{itemize}
	
	\subsection{Influence of $\text{M}_{n1}, \text{M}_{d1}, \text{M}_{n2}$ and $\text{M}_{d2}$}
	
		\begin{table}[htb!]
		\tbl{Number of initial conditions converging towards different libration points and the values of  entropy $\text{S}_{b}$ and $\text{S}_{bb}$ due to variation in parameters $\text{M}_{n1}, \text{M}_{d1}, \text{M}_{n2}$ and $\text{M}_{d2}$. $R$=2 is fixed.} 
		{\begin{tabular}{|c | c| c| c| c| c| c| p{1.5 cm}| c| c|}
				\hline
				{$\text{M}_{n1}$, $\text{M}_{d1}$, $\text{M}_{n2}$, $\text{M}_{d2}$} &$\text{Total Points}$&$ \text{L1}$ & $ \text{L2}$ & $ \text{L3}$ & $ \text{L4}$ & $ \text{L5}$&Execution Time &$\text{S}_{b}$&$\text{S}_{bb}$ \\[2.5pt]\hline
				{0.08, 2.0, 0.04, 0.6} &1012300& 476105&104226& 19413& 206381& 206195&2368.86& 0.806377&0.930766 \\[1pt]\hline
				{0.06, 1.8, 0.06, 0.8} &1043359&477939&93673&28561&221538&221648&2628.81&0.941639&1.06536  \\[1pt]\hline
				{0.04, 1.6, 0.08, 1.0}&1073073&595431&63702&30089&191832&192019&1812.15&0.861113&0.985576 \\[1pt]\hline
				{0.02, 1.4, 0.1, 1.2}&1037341&532105&46465&42676&208134&207961&2552.17&0.661595&0.882299  \\[1pt]\hline
				{0.01, 1.2, 0.12, 1.4}&1011026&482757&37538&60105&215528&215098&3539.01&0.970159&1.05315  \\[1pt]\hline
				{0.008, 1.0, 0.14, 1.6}&1020700& 527815&27792& 72293& 196344& 196456&2429.71&0.948648&1.06536  \\[0pt]
				\hline
		\end{tabular}} 
	\end{table}
	
	\begin{figure*}[htb!]
		\centering
		\begin{tabular}{cc}
			\includegraphics[scale=.4]{bq21}~~~~~~~~~ &
			\includegraphics[scale=.4]{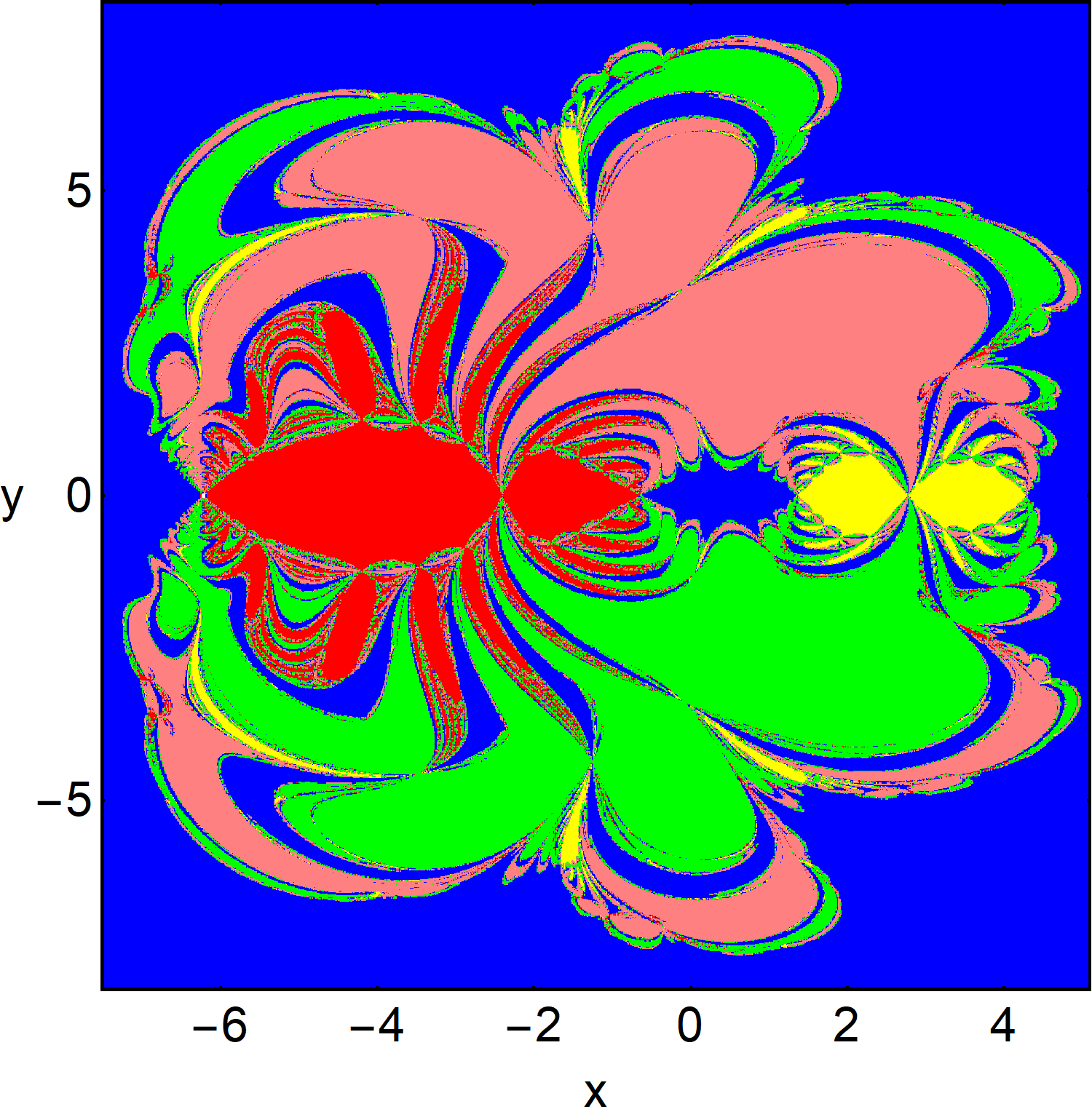}\\
			(a)  &(b)  \\
			{}&{}\\
			\includegraphics[scale=.4]{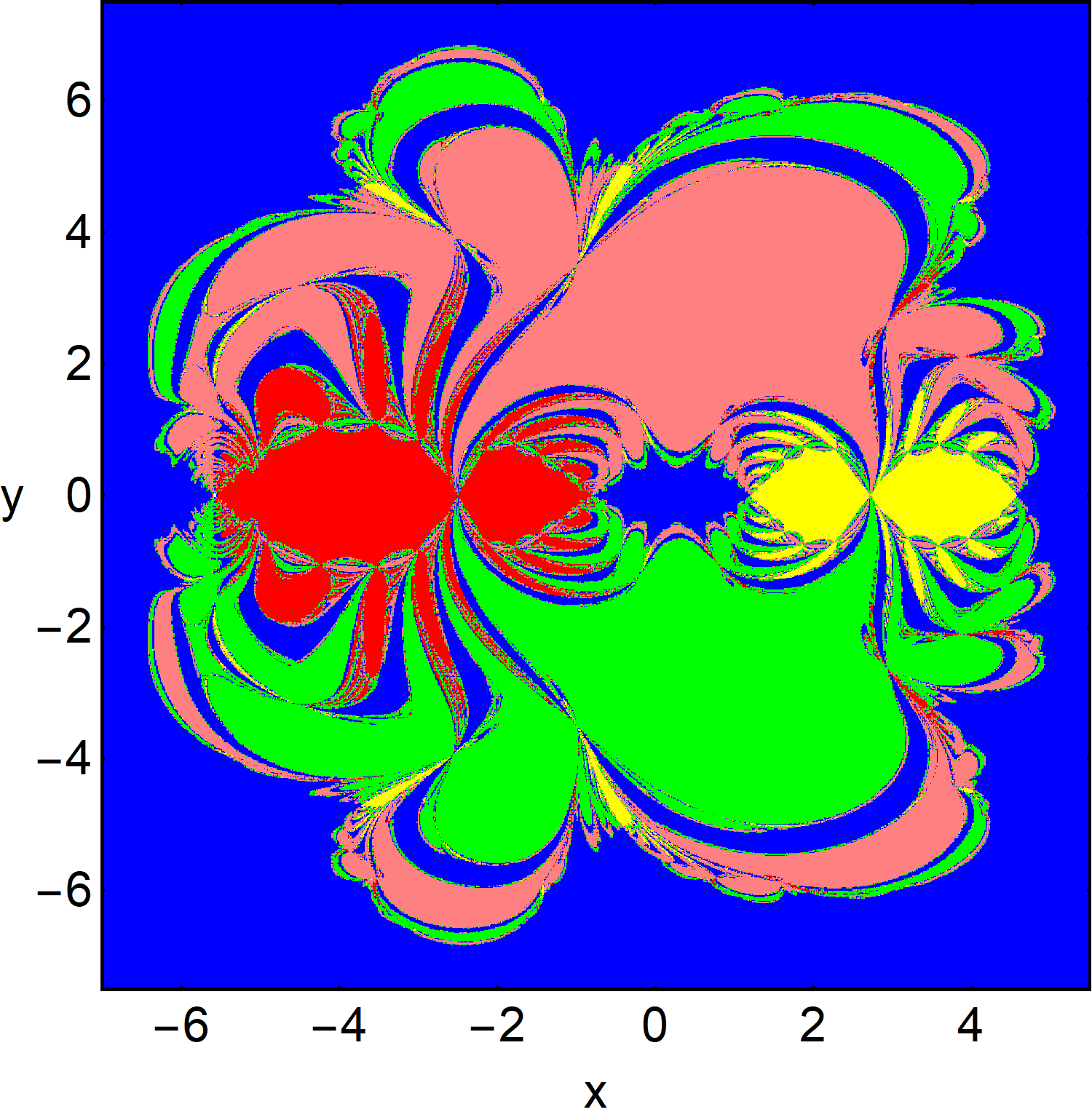}~~~~~~~~~~&
			\includegraphics[scale=.4]{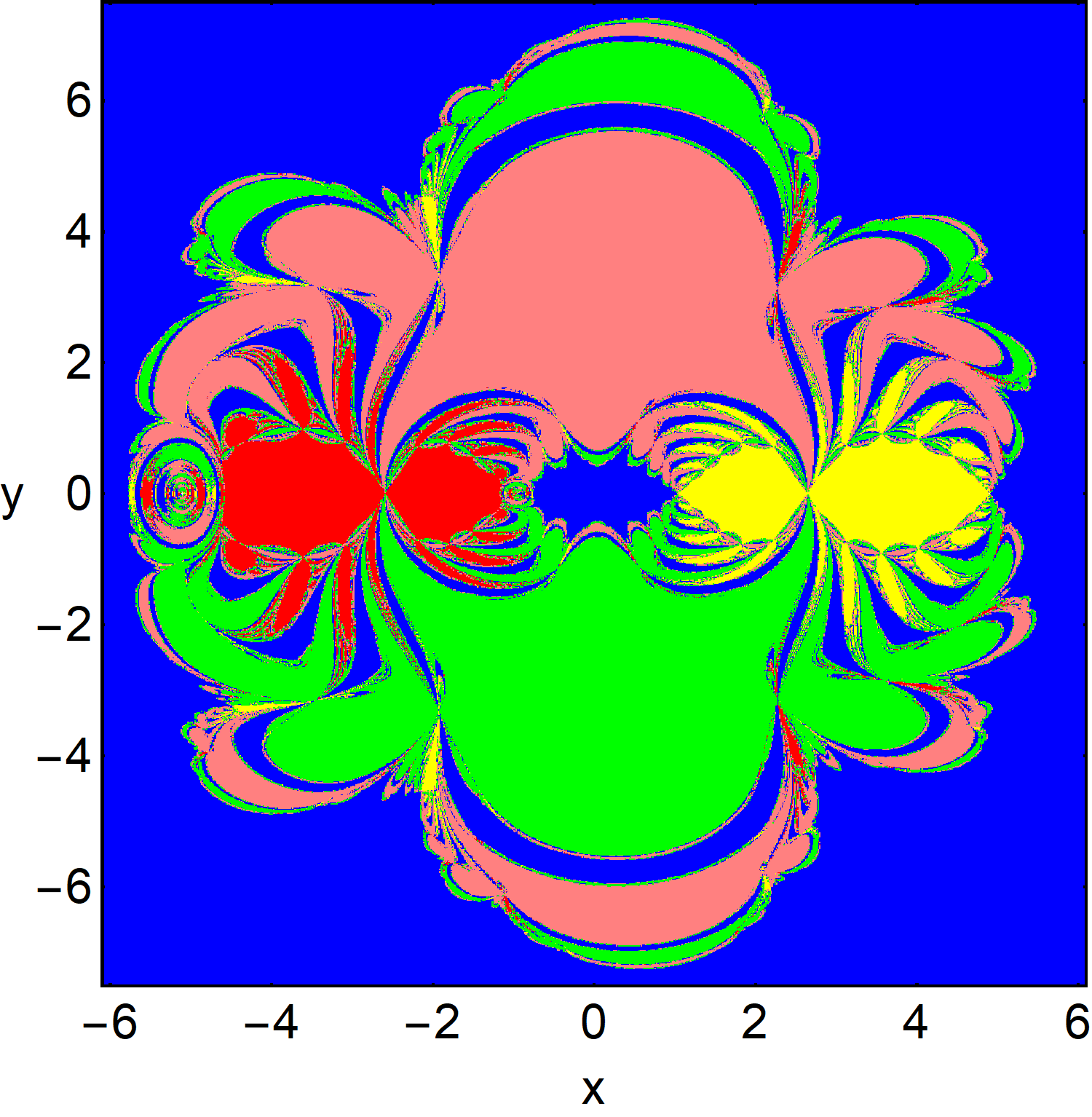}\\		
			(c) &(d) \\
			{}&{}\\
			\includegraphics[scale=.4]{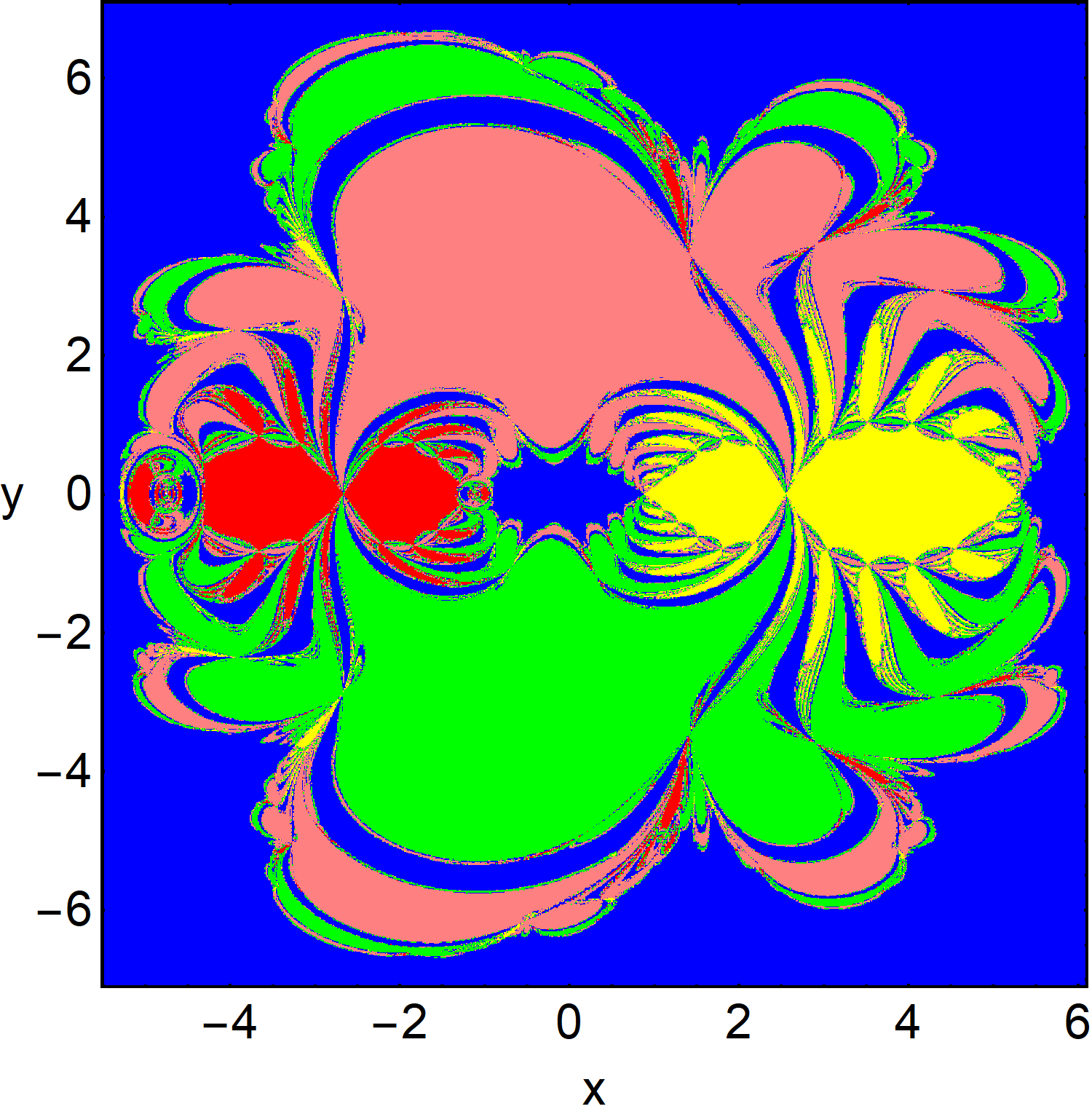}~~~~~~~~~~&
			\includegraphics[scale=.4]{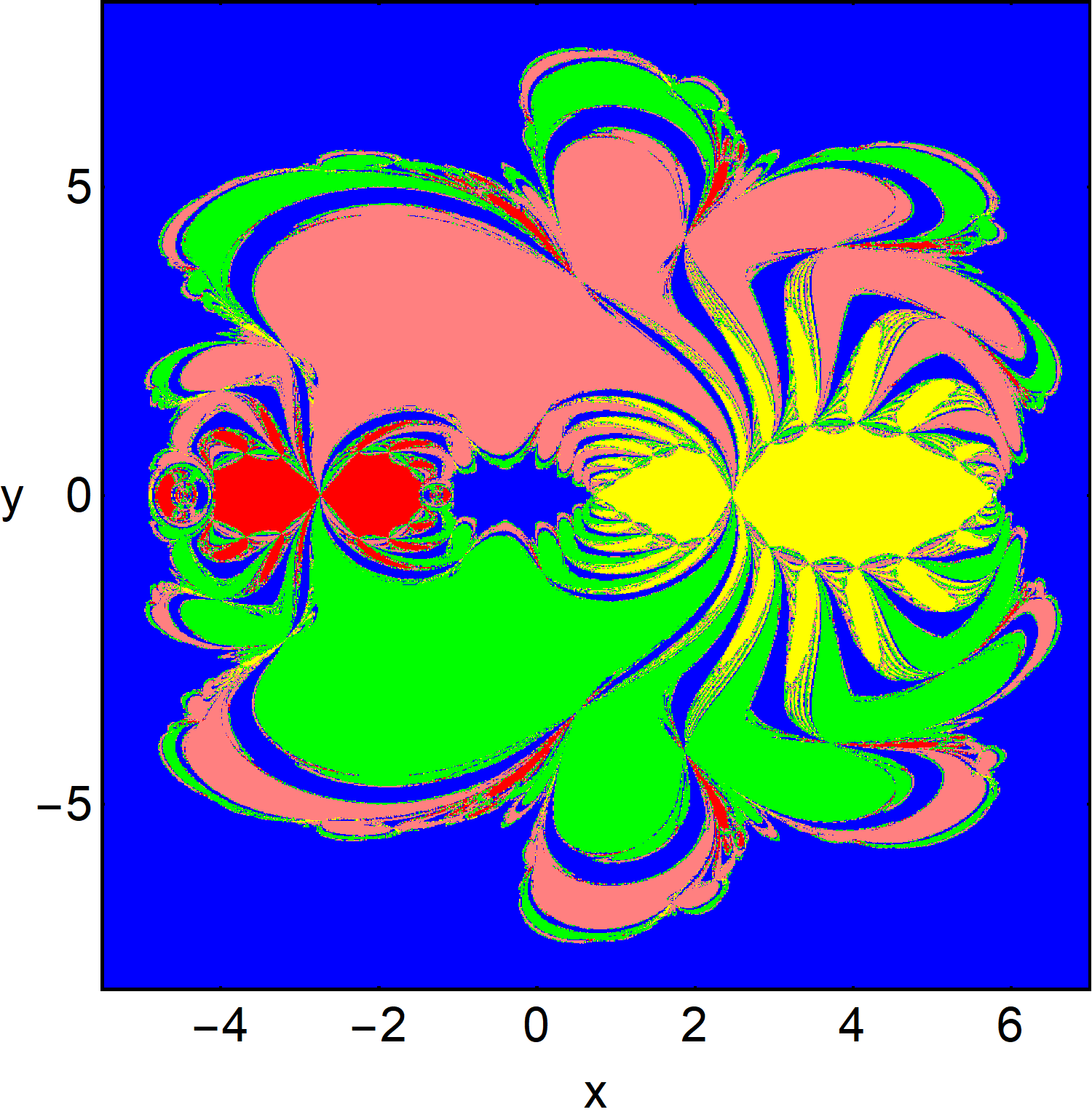}\\		
			(e) &(f) \\
		\end{tabular}
		\caption{BoA for different values of $\text{M}_{n1}$, $\text{M}_{d1}$, $\text{M}_{n2}$ and $\text{M}_{d2}$. The colour codes for the BoA corresponding to libration points are $\text{L}1$ (blue), $\text{L}2$ (Red), $\text{L}3$ (Yellow) , $\text{L}4$ (Pink) and $\text{L}5$ (Green). The different values of the parameters are (a) $\text{M}_{n1}$ = 0.08; $\text{M}_{d1}$ = 2.0; $\text{M}_{n2}$ = 0.04; $\text{M}_{d2}$ = 0.6;(b) $\text{M}_{n1}$ = 0.06; $\text{M}_{d1}$ = 1.8; $\text{M}_{n2}$ = 0.06; $\text{M}_{d2}$ = 0.8;(c) $\text{M}_{n1}$ = 0.04; $\text{M}_{d1}$ = 1.6; $\text{M}_{n2}$ = 0.08; $\text{M}_{d2}$ = 1;(d) $\text{M}_{n1}$ = 0.02; $\text{M}_{d1}$ = 1.4; $\text{M}_{n2}$ = 0.1; $\text{M}_{d2}$ = 1.2;(e) $\text{M}_{n1}$ = 0.01; $\text{M}_{d1}$ = 1.2; $\text{M}_{n2}$ = 0.12; $\text{M}_{d2}$ = 1.4;(f) $\text{M}_{n1}$ = 0.008; $\text{M}_{d1}$ = 1.0; $\text{M}_{n2}$ = 0.14; $\text{M}_{d2}$ = 1.6. $R$=2 is fixed.}
	\end{figure*}
	
	\begin{figure*}[htb!]
		\centering
		\begin{tabular}{cc}
			\includegraphics[scale=.5]{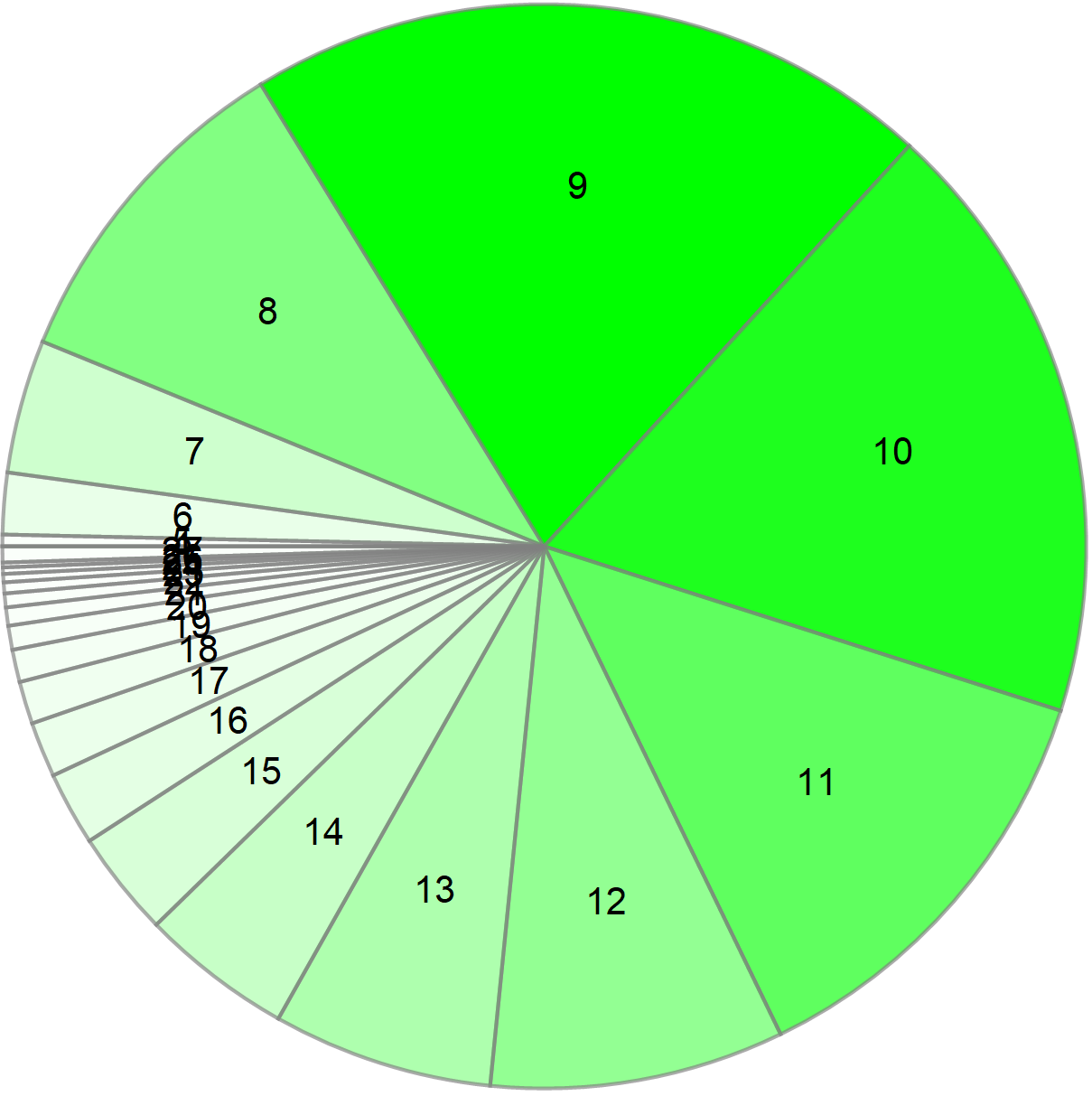} ~~~~~~~~&
			\includegraphics[scale=.5]{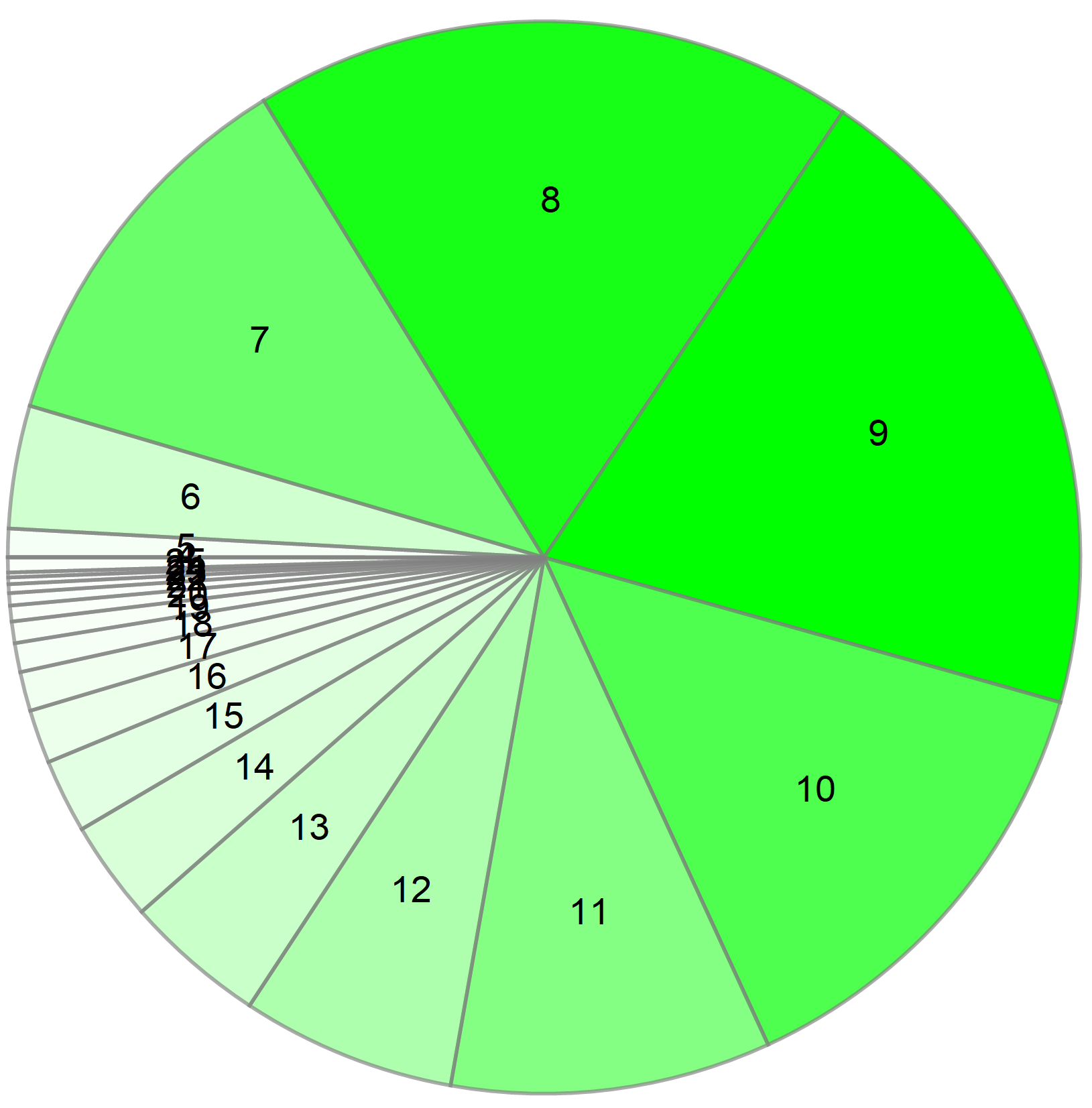}\\
			(a) MINo.=9 ~~~~~~~&(b) MINo.=9 \\
			{}&{}\\
			\includegraphics[scale=.5]{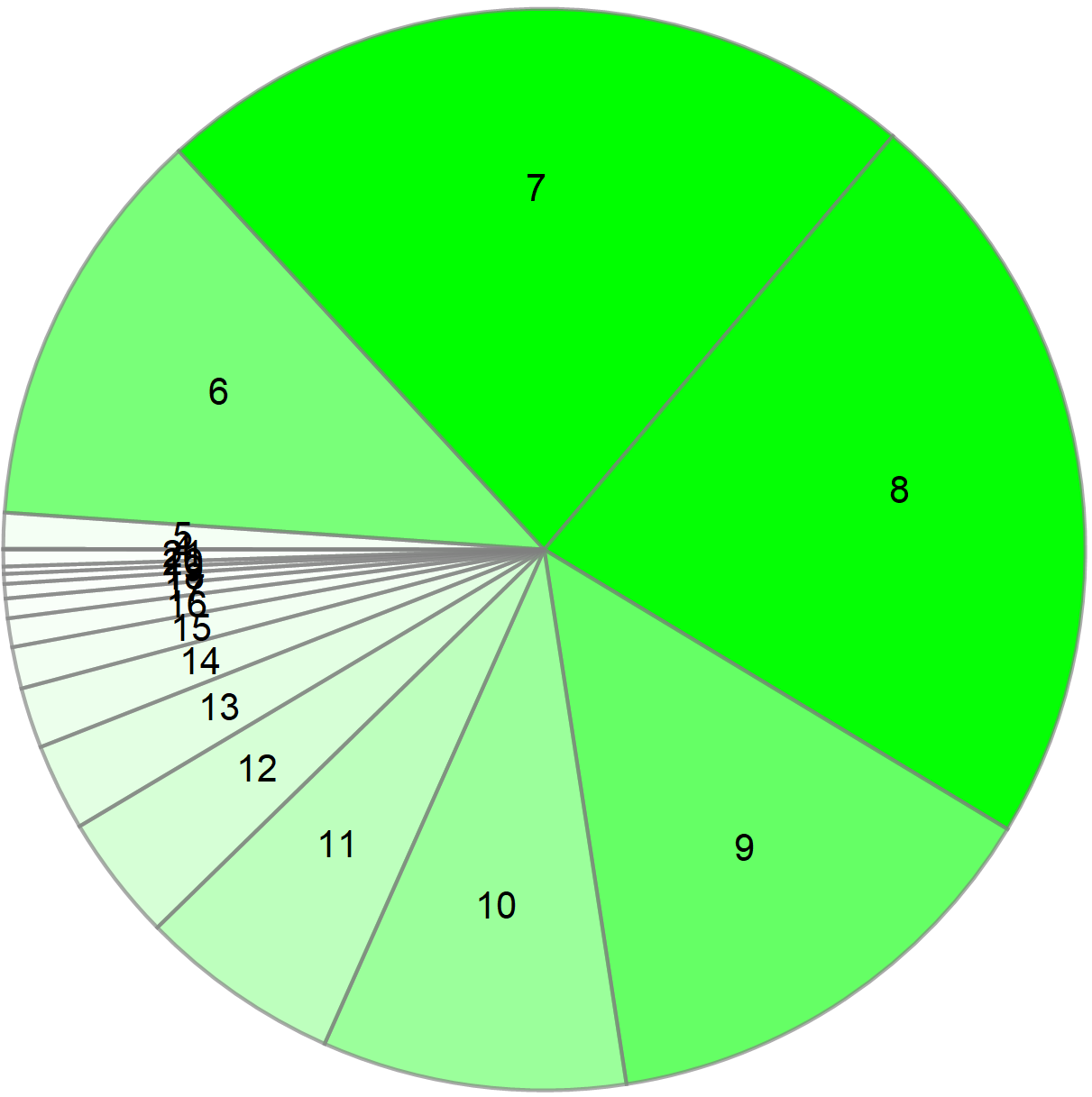}~~~~~~~~&
			\includegraphics[scale=.5]{bq253pi}\\		
			(c) MINo.=7~~~~~~~&(d)MINo.=7\\
			{}&{}\\
			\includegraphics[scale=.5]{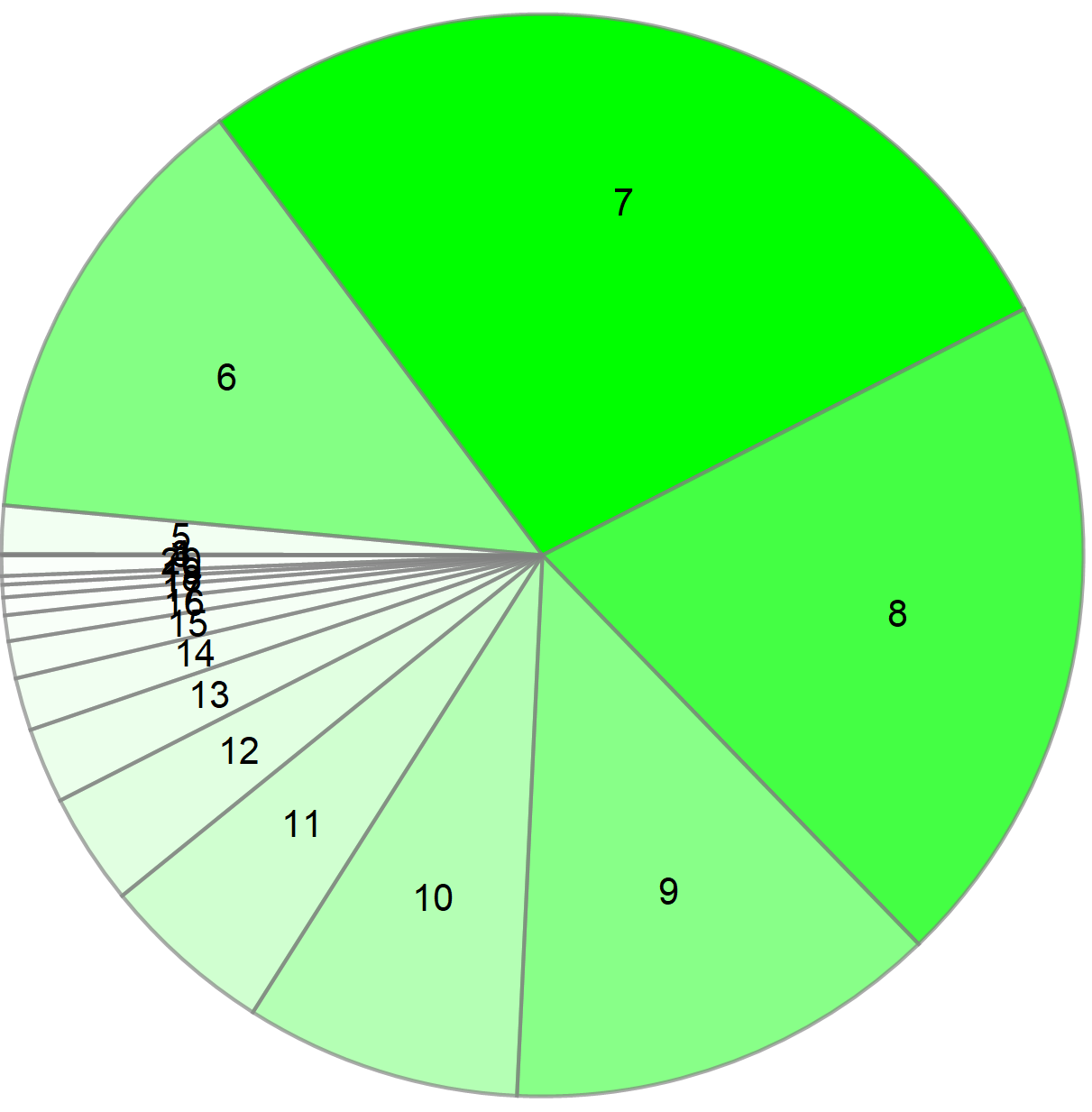}~~~~~~~~~&
			\includegraphics[scale=.5]{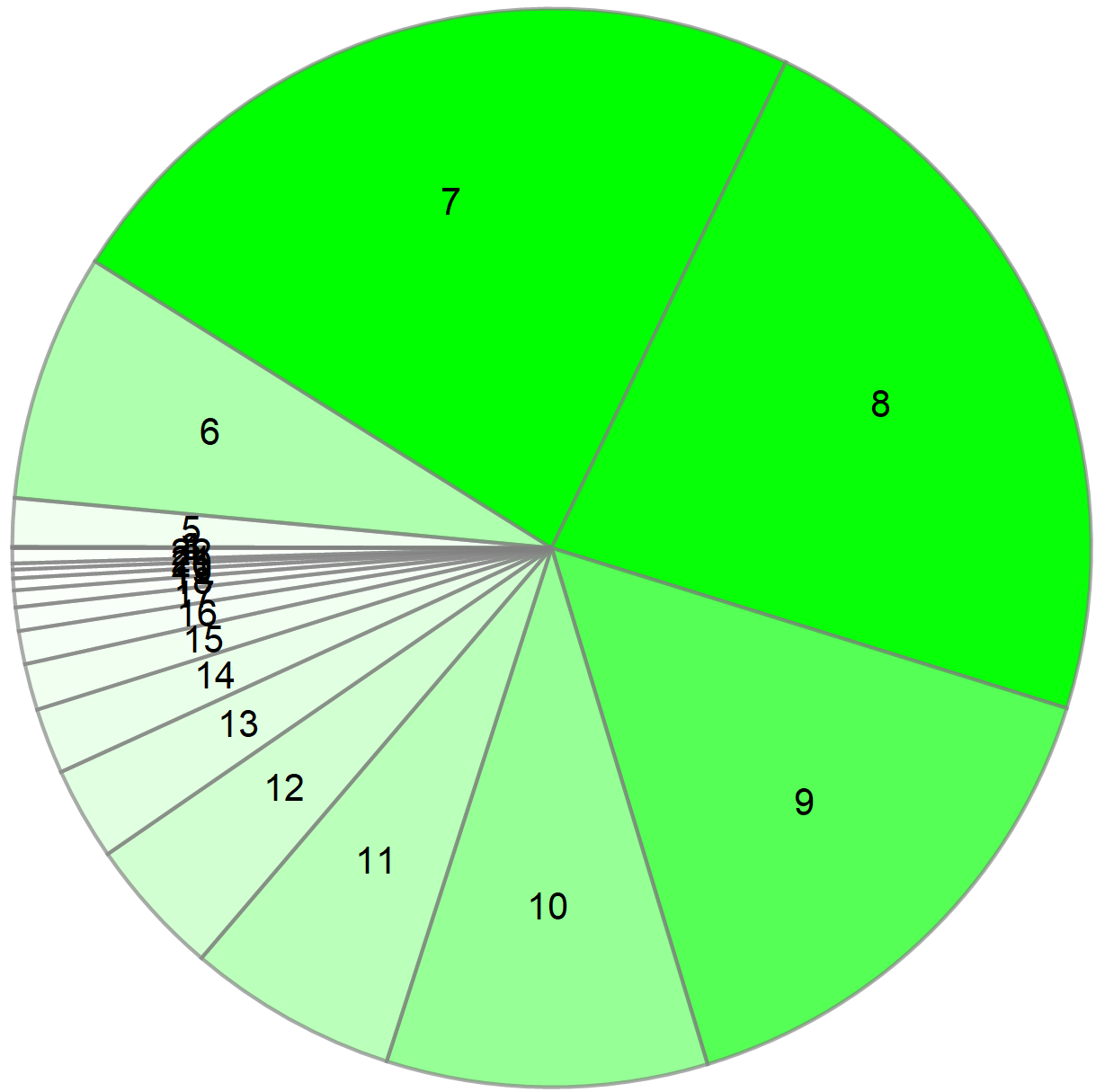}\\		
			(e) MINo.=7~~~~~~~~&(f) MINo.=8\\
		\end{tabular}
		\caption{(a-f) $\text{Pie-chart}$ of number of iterations taken ($\text{N}$) for different values of $\text{M}_{n1}$, $\text{M}_{d1}$, $\text{M}_{n2}$ and $\text{M}_{d2}$. The different values of the parameters are (a) $\text{M}_{n1}$ = 0.08; $\text{M}_{d1}$ = 2.0; $\text{M}_{n2}$ = 0.04; $\text{M}_{d2}$ = 0.6;(b) $\text{M}_{n1}$ = 0.06; $\text{M}_{d1}$ = 1.8; $\text{M}_{n2}$ = 0.06; $\text{M}_{d2}$ = 0.8;(c) $\text{M}_{n1}$ = 0.04; $\text{M}_{d1}$ = 1.6; $\text{M}_{n2}$ = 0.08; $\text{M}_{d2}$ = 1;(d) $\text{M}_{n1}$ = 0.02; $\text{M}_{d1}$ = 1.4; $\text{M}_{n2}$ = 0.1; $\text{M}_{d2}$ = 1.2;(e) $\text{M}_{n1}$ = 0.01; $\text{M}_{d1}$ = 1.2; $\text{M}_{n2}$ = 0.12; $\text{M}_{d2}$ = 1.4;(f) $\text{M}_{n1}$ = 0.008; $\text{M}_{d1}$ = 1.0; $\text{M}_{n2}$ = 0.14; $\text{M}_{d2}$ = 1.6. $R$=2 is fixed. MINo. denotes the maximum number of  iterations.}
	\end{figure*}
	
	\begin{figure*}[htb!]
		\centering
		\begin{tabular}{cc}
			\includegraphics[scale=.27]{bq23}~~~~~ &
			\includegraphics[scale=.32]{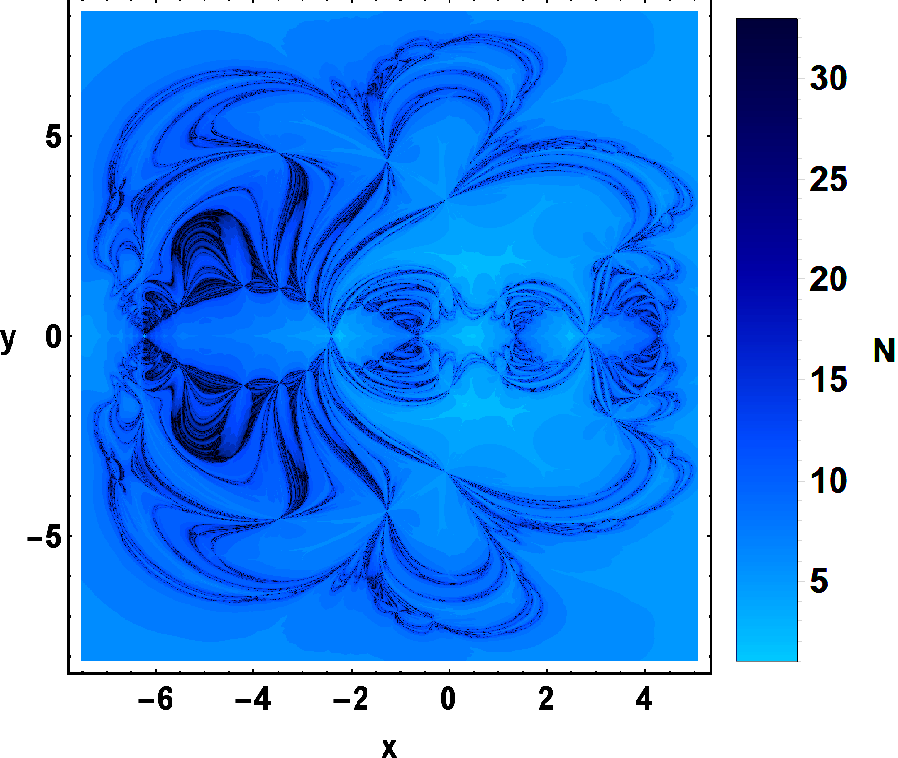}\\
			(a) &(b)  \\
			{}&{}\\
			\includegraphics[scale=.4]{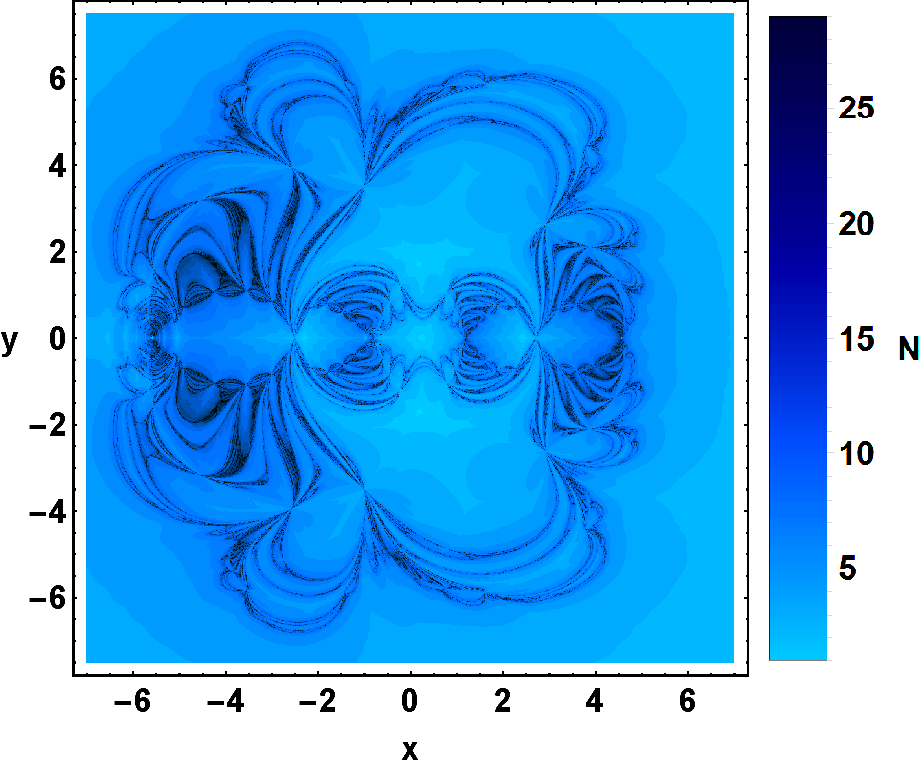}~~~~~&
			\includegraphics[scale=.4]{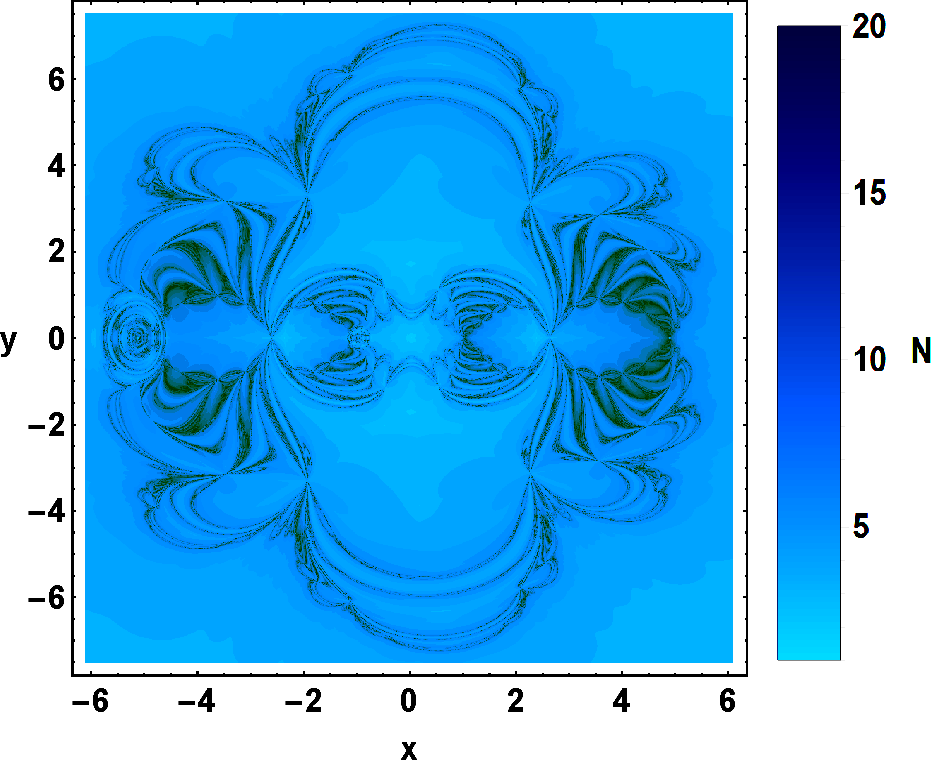}\\		
			(c)  &(d) \\
			{}&{}\\
			\includegraphics[scale=.4]{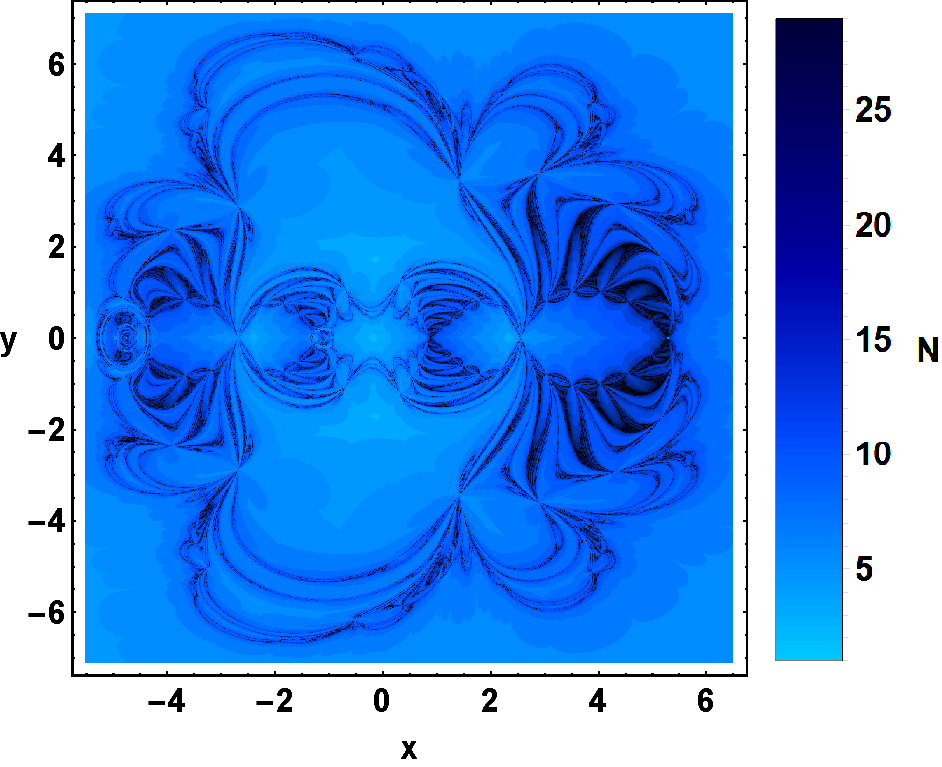}~~~~~&
			\includegraphics[scale=.4]{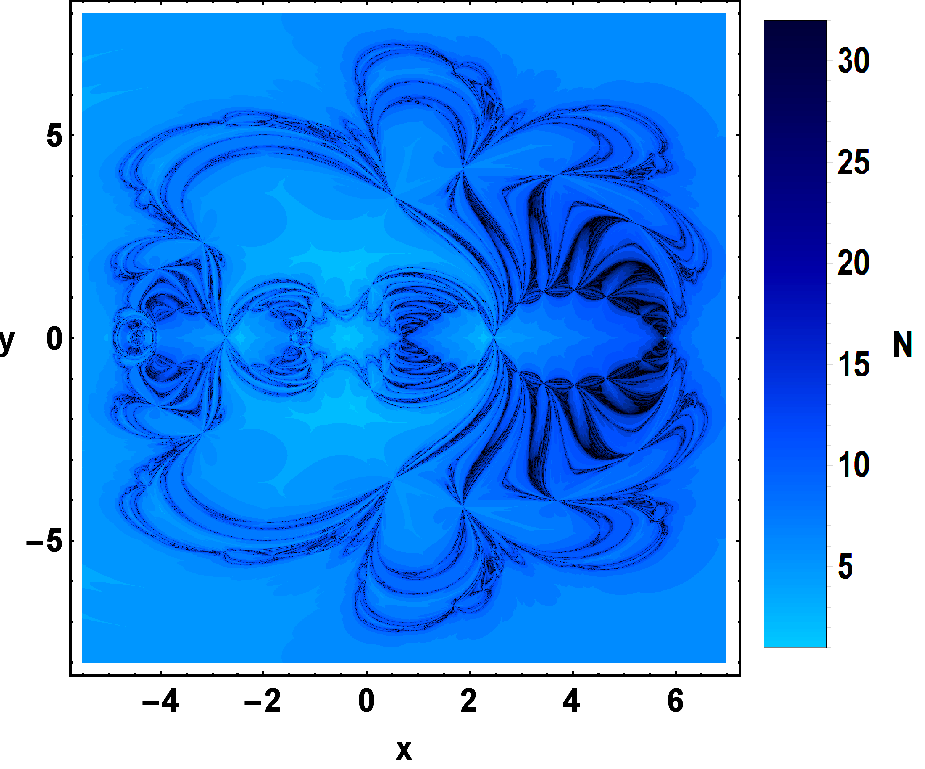}\\		
			(e) &(f) \\
		\end{tabular}
		\caption{(a-f) Iterations \textbf{N} (blue tone) needed to achieve desired accuracy for different values of $\text{M}_{n1}$, $\text{M}_{d1}$, $\text{M}_{n2}$ and $\text{M}_{d2}$. The different values of the parameters are (a) $\text{M}_{n1}$ = 0.08; $\text{M}_{d1}$ = 2.0; $\text{M}_{n2}$ = 0.04; $\text{M}_{d2}$ = 0.6;(b) $\text{M}_{n1}$ = 0.06; $\text{M}_{d1}$ = 1.8; $\text{M}_{n2}$ = 0.06; $\text{M}_{d2}$ = 0.8;(c) $\text{M}_{n1}$ = 0.04; $\text{M}_{d1}$ = 1.6; $\text{M}_{n2}$ = 0.08; $\text{M}_{d2}$ = 1;(d) $\text{M}_{n1}$ = 0.02; $\text{M}_{d1}$ = 1.4; $\text{M}_{n2}$ = 0.1; $\text{M}_{d2}$ = 1.2;(e) $\text{M}_{n1}$ = 0.01; $\text{M}_{d1}$ = 1.2; $\text{M}_{n2}$ = 0.12; $\text{M}_{d2}$ = 1.4;(f) $\text{M}_{n1}$ = 0.008; $\text{M}_{d1}$ = 1.0; $\text{M}_{n2}$ = 0.14; $\text{M}_{d2}$ = 1.6. $R$=2 is fixed.}
	\end{figure*}

	We continue our investigations with another case where the masses of a nucleus $\text{M}_{n1}$ and the disk $\text{M}_{d1}$of a galaxy (G1) is decreasing and the masses of nucleus $\text{M}_{n2}$ and the disk $\text{M}_{d2}$ of galaxy G2 is increasing. We have displayed BoA for six values in Fig. 6(a-f). In all cases, the value of $\text{R}=2$ is fixed. In Fig. 7(a-f), the Pie-chart related to different parameters is displayed. Fig.8(a-f) comprises of the different graphs representing the iterations needed for each initial conditions.
	
	Further, in Fig. 6(a-f), we notice that the BoA for $\text{L2}$ is decreasing, and for $\text{L3}$, it is increasing due to variation in parameters. We observe in Fig. 6(a) few small regions in the left part which disappeared in the next Fig. 6(b) due to change in parameters. We notice such changes or other when we move from Fig. 6(a) to Fig. 6(f). Similar to the previous case, the BoA corresponding to $\text{L1}$ extends to infinity. The BoA of $\text{L4}$ and $\text{L5}$ remain more or less the same in all cases. The shape of BoA of libration points $\text{L4}$ and $\text{L5}$ looks like many butterfly wings. We also observe that the BoA of libration points $\text{L2}$ and $\text{L3}$ have the shape like insects with legs and antennas. 
	
	Similar to Table 1, Table 2 comprises the number of initial conditions converging towards different libration points. The value of basin entropy and boundary basin entropy is also given in Table 2. By Table 2, it is evident that the almost entire region in the BoA (except one case) is fractal. The value of $\text{S}_{\text{bb}}$ is more than the value of $\text{S}_{\text{b}}$ indicates the presence of more fractal regions along the boundaries. 
	
	The number of iterations needed for convergence towards libration points for all initial conditions varies from 5 to 30. The distribution of iterations is shown using Pie-chart in Fig. 7 (a-f). The maximum probable iterations are 7 or 8 or 9 in almost all cases.  Similar to the case I, 95\% initial conditions converge in 25 iterations. The number of iteration needed to convergence for each initial condition is displayed in Fig. 8(a-f).  The number of required iterations is presented using a blue tone, i.e., deep blue tone represents more iteration than faded blue tone. This figure gives us a clear idea of the regions where there is a need for more number of required iterations. We summaries followings from Table 2, Fig. 6, Fig. 7 and Fig. 8.

	\begin{itemlist}
		\item The total area covered by BoA corresponding to $\text{L2, L3, L4}$ and $\text{L5}$ are finite whereas the area covered by a BoA of libration point $\text{L1}$ extends to infinity. The configuration plane $(x, y)$ is filled by BoA completely.
		\item All initial conditions are converging towards one of the five libration points with the accuracy $10^{-15}$. We do not find any non-converging points. 
		\item Except for one value of a parameter, the existence of fractal is verified by the values of basin entropy given in Table 2. However, in this particular case, the boundaries of BoA are fractal. In all other cases, there is a presence of highly fractal regions along the boundaries (See Table 2 and Fig. 8(a-f)). 
		
		\item In all cases, the maximum number of iterations needed for the convergence of $95 \%$ of the initial conditions are below 15. The maximum probable number of iterations needed to achieve the desired accuracy lies between 7 to10. 
		
	\end{itemlist}  
	
	\section{Concluding Remarks}
The novelty and importance of this work are clear from the fact that we have chosen this model for the first time to study the BoA and the measure of uncertainty involved in it. Besides this, we have investigated the existence of fractal regions in BoA. The binary system of interacting galaxies has been explored in the framework of the Circular Restricted Three-Body Problem. The parametric evolution of libration points is shown in Fig. 1. The impact of Jacobi's constant on the geometry of zero-velocity curves is shown in Fig. 2. The multivariate form of the NR method is used to find the convergence of initial conditions towards libration points (act as attractors in the present case). With the help of shape of basins, the influence of parameters $R$ and  $\text{M}_{n1}, \text{M}_{d1}, \text{M}_{n2}$, $\text{M}_{d2}$ is analysed (Fig. 3 and 5). The data obtained from numerical simulations is given in Table 1 and Table 2, which reflects the originality of this work. Table 1 and Table 2 comprises of the details convergence of initial conditions towards libration points; time consumed by CPU; the value of basin entropy and boundary basin entropy. The number of iterations needed for the convergence of the initial conditions is shown using Pie-chart (in the tone of green colour) in Fig. 4 and Fig. 7, which is useful, informative and different from earlier works. Further, we have established the relationship between the number of iterations required for convergence and the set of initial conditions on configuration plane $(x,y)$ (Fig. 5 and 8). 

For all numerical simulations, we have used a machine configured with Intel(R) Core(TM)  i7-8550U CPU 1.80 GHz. In all cases, the computational time of the CPU is less than or equal to one hour for the simulation a uniform grid of 1024$\times$1024 initial conditions (nearly ten lakhs). The time taken by the CPU to classify all initial conditions is less than one minute. We have considered six different values of all the parameters. The results of simulations can be summarized as follows:
\begin{itemlist}
	
	\item The programming for the computation of BoA; basin entropy and plotting of all graphs is done on Mathematica. The execution time taken by CPU for BoA and the classification of initial conditions in this model is explicitly mentioned in Table 1 and Table 2. As per data available from earlier works, the time mentioned in Table 1 and Table 2 is comparatively less.    
	
	\item In almost all cases, we find that configuration plane $(x,y)$ contains the combinations of fractal regions and distinct BoA. If we choose an initial condition in these regions, it is challenging to predict the convergence towards a specific libration point. The unpredictability of basin of attraction is due to the non-linearity of the expressions of the potential responsible for the motion of a star in the presence of the two interacting galaxies (G1 and G2) (See equation (4)) ).  These regions are mainly located along with the neighbourhood of boundaries of BoA.
	
	\item The area enclosed by BoA corresponding to libration points $\text{L2, L3, L4 and L5}$ are finite whereas, the area associated with central libration point $\text{L1}$ extends to infinity. Also, the area of basins corresponding to $\text{L4}$ and $\text{L5}$ are approximately the same for all cases. The area of basins corresponding to $\text{L2}$ and $\text{L3}$ changes due to variation in the parameters.
	
	\item Based on simulations, we observe that all the initial conditions (10 lakhs approx.) on a uniform grid of 1024$\times$1024 converge to any one of the five libration points of the dynamical system with an accuracy of order $10^{-15}$. We do not find any non-converging initial condition. (See Table 1 and Table 2)
	
	\item Fig. 5 (a-f) and Fig. 8 (a-f)  reveal that all initial conditions which require more iterations lie along the boundaries of BoA, i.e., along with the fractal regions as compare to Initial conditions lying away from boundaries.
	
	\item Due to a decrease in the mass of the galaxy (\text{G1}) and the increase of the mass of the galaxy (\text{G2}), the area of a BoA due to $\text{L2}$ decreases and the area of a BoA due to $\text{L3}$ increases. However, the region occupied by BoA due to $\text{L4}$ and $\text{L5}$ is roughly the same for all cases. (See Fig. 6 (a-f))
	
	\item By Table 1 and Table 2, we find that the values of entropy $\text{S}_{b}$ and $\text{S}_{bb}$ is greater than $\log{2}$ except for one value of $\text{S}_{b}$ in Table 2. Thus all BoA are fractal and there is an existence of highly fractal regions along the boundaries.
	
	\item The maximum number of iteration that an initial condition needs to converge towards libration points lie between 7-10 for all cases. (See Fig. 4 (a-f)and Fig. 7 (a-f)). 
	
	\item  The parameters have a significant impact on the evolution of libration points as well as the geometry of zero-velocity curves.

\end{itemlist}
We observe very few research works in the area of space dynamics. That is why we have considered a galactic model for our work. These results will be useful for many researchers to work in these models. On the other hand,  these observations give us an intuitive idea for the existence of some properties (like a different type of basins, the existence of Wada basin ). In future, we will try to investigate these properties in other complex nonlinear models in the field of space dynamics and celestial mechanics.

 	\end{document}